\documentclass[DIV=15, bibliography=totoc]{scrartcl}

\usepackage[a4paper,margin=2.25cm,bottom=2.5cm]{geometry}
\usepackage{float}
\usepackage[export]{adjustbox}
\usepackage[normalem]{ulem}
\usepackage{diagbox}
\usepackage{stmaryrd}

\usepackage{xcolor} 
\definecolor{lightblue}{rgb}{0,0.5,1.0}
\definecolor{linkblue}{rgb}{0,0.1,0.6}
\definecolor{citegreen}{rgb}{0,0.4,0.0}
\definecolor{linkred}{rgb}{0.8,0,0.005}
\definecolor{mailviolet}{rgb}{0.3,0,0.35}
\definecolor{tumblue}{rgb}{0,0.396,0.741}
\definecolor{darkgreen}{rgb}{0,0.4,0} 
\definecolor{darkbrown}{rgb}{0.5, 0.396, 0.09}

\usepackage[hyperindex,colorlinks=true,linkcolor=linkblue,citecolor=citegreen,urlcolor=mailviolet,filecolor=linkred]{hyperref}
\usepackage[hyphenbreaks]{breakurl}
\usepackage{caption, subcaption}
\usepackage{booktabs}   
\usepackage{amsmath}
\usepackage{amsfonts}
\usepackage{lmodern}
\usepackage{graphicx}
\usepackage{listings}
\usepackage{todonotes}
\usepackage{placeins}
\usepackage{fancyhdr}
\usepackage{wrapfig}

\usepackage{siunitx}
\usepackage[square,comma,nonamebreak,compress,numbers]{natbib}

\usepackage{soulpos}
\usepackage{ulem}
\usepackage{authblk} 

\usepackage{pgfplots}
\usepackage{pgfplotstable}
\pgfplotsset{compat = newest}
\usepackage{filecontents}
\usepackage{tikz}
\usepackage{tikzscale}
\usepackage{tikz-3dplot}

\usetikzlibrary{spy}
\usetikzlibrary{intersections}
\usetikzlibrary{arrows,shapes}
\usetikzlibrary{spy}
\usetikzlibrary{backgrounds}  
\usetikzlibrary{decorations}
\usetikzlibrary{decorations.markings}
\usetikzlibrary{positioning}
\usetikzlibrary{patterns}
\usetikzlibrary{calc} 
\usetikzlibrary{quotes} 
\usetikzlibrary{external} 
\usetikzlibrary{matrix}

\pgfplotscreateplotcyclelist{customCycleList}
{%
  {blue, mark=o},
  {red, mark=square},
  {darkgreen, mark=diamond},
  {cyan, mark=diamond},
  {darkbrown, mark=triangle*},
  {black, mark=pentagon},
}

\pgfplotsset{every axis/.append style= {
    cycle list name=customCycleList,
}}

\title{Isogeometric Multi-Resolution Full Waveform Inversion based on the Finite Cell Method}

\author[1]{Tim B\"urchner\thanks{\href{mailto:tim.buerchner@tum.de}{\texttt{tim.buerchner@tum.de}},
		Corresponding author}}
\author[1]{Philipp Kopp}
\author[1]{Stefan Kollmannsberger}
\author[1, 2]{Ernst Rank}

\affil[1]{Chair of Computational Modeling and Simulation, 
	Technische Universit\"at M\"unchen 
}

\affil[2]{Institute for Advanced Study, 
	Technische Universit\"at M\"unchen 
}

\date{}
\fancypagestyle{plain}{%
	\fancyhf{}%
	\fancyfoot[L]{
		\vspace{-0.0cm} 
		\footnotesize{Preprint submitted to \textit{Computer Methods in Applied Mechanics and Engineering}.}  
} }

\newcommand{\tensor}[1]{\mathbf{#1}}
\newcommand{\tabref}[1]{Table~\ref{#1}}
\newcommand{\figref}[1]{Figure~\ref{#1}}
\newcommand{\figsref}[1]{Figures~\ref{#1}}
\newcommand{\secref}[1]{Section~\ref{#1}}

\usepackage{xcolor}
\usepackage{comment} 

\begin{document}

\normalem \maketitle
\normalfont\fontsize{11}{13}\selectfont

\definecolor{mygreen}{rgb}{0.0, 0.5, 0.0}


\vspace{-1.5cm} \hrule 

\section*{Abstract}

Full waveform inversion (FWI) is an iterative identification process that serves to minimize the misfit of model-based simulated and experimentally measured wave field data, with the goal of identifying a field of parameters for a given physical object. For many years, FWI has been used very successfully in seismic imaging to deduce velocity models of the earth or of local geophysical exploration areas. FWI has also been successfully applied in various other fields, including non-destructive testing (NDT) and biomedical imaging. The inverse optimization process of FWI is based on forward and backward solutions of the (elastic or acoustic) wave equation, as well as on efficient computation of an adequate optimization direction. Many approaches use (low order) finite element or finite difference methods, often with a field of parameter values with a resolution corresponding to elements or nodes of the discretized wave field. In a previous paper \cite{Buerchner2023}, we explored opportunities of using the finite cell method (FCM) as the wave field solver, which has the advantage that highly complex geometric models can be incorporated easily. Furthermore, we demonstrated that the identification of the model’s density outperforms that of the velocity -- particularly in cases where unknown voids characterized by homogeneous Neumann boundary conditions need to be detected. The paper at hand extends this previous study in the following aspects: The isogeometric finite cell analysis (IGA-FCM) -- a combination of isogeometric analysis (IGA) and FCM -- is applied for the wave field solver, with the advantage that the polynomial degree and subsequently also the sampling frequency of the wave field can be increased quite easily. Since the inversion efficiency strongly depends on the accuracy of the forward and backward wave field solution and of the gradient of the functional, consistent and lumped mass matrix discretization are compared. The resolution of the grid describing the unknown material density -- thus allowing to identify voids in a physical object -- is then decoupled from the knot span grid. Finally, we propose an adaptive multi-resolution algorithm that refines the material grid only locally using an image processing-based refinement indicator. The developed inversion framework allows fast and memory-efficient wave simulation and object identification. While we study the general behavior of the proposed approach on 2D benchmark problems, a final 3D problem shows that it can also be used to identify void regions in geometrically complex spatial structures.

\vspace{0.25cm}
\noindent \textit{Keywords:} 
full waveform inversion, isogeometric analysis, finite cell method, multi-resolution, scalar wave equation
\vspace{0.25cm}


\section{Introduction}
\FloatBarrier
{
Tom Hughes has initiated and driven countless innovations in computational science and engineering. Among the most important is undoubtedly the invention of isogeometric analysis. Originally motivated by the goal of uniting the separate worlds of geometric modeling and finite element analysis, the great value of this method is also demonstrated by the fact that new areas of application continue to emerge that were not part of the original objective. The present contribution is exactly of this kind. We show how unknown geometric features of a structure can be effectively identified, and how an inverse analysis benefits from superior inherent properties of IGA, such as the ability to obtain highly accurate results with a small number of degrees of freedom. \\ Happy Birthday, Tom.

With its origins in the 1980s~\cite{Lailly1983, Tarantola1984}, full waveform inversion~(FWI) has become a well-established method in the field of seismic tomography. Waves traveling through the interior of a medium are measured and compared to model-based simulated wave signals. Information about internal material properties is extracted in a nonlinear optimization problem. The adjoint method using forward and backward simulations of the wave field allows for an efficient use of gradient-based optimization~\cite{Givoli2021}. A comprehensive introduction to FWI can be found in~\cite{Fichtner2011}, reviews in~\cite{Virieux2009, Vigh2008}.
	

While FWI has been used very successfully in geophysics for a long time, its application to biomedical applications~\cite{Pratt2007, Sandhu2017, Guasch2020} and non-destructive testing~(NDT)~\cite{Seidl2018, Rao2017, Lin2020} has gained traction in recent years. In these problems, the goal often is to identify interior voids or fractures characterized by homogeneous Neumann boundary conditions. Such defects are challenging to detect using classical velocity-based FWI. In previous work~\cite{Buerchner2023}, we showed that density inversion allows to efficiently identify and reconstruct these defects in possibly damaged samples. A thorough analysis of the distinct behaviors of density and velocity inversion based on a boundary layer description is provided in~\cite{Rabinovich2023}. To ensure the success of FWI, it is crucial to efficiently and precisely solve the forward and backward wave problems -- and to construct an accurate and sufficiently resolved material description. However, many numerical schemes couple the wave field discretization and material representation, which does not allow to freely adapt them independently of each other. It is common to use a finite difference grid or a finite element mesh for the wave field to describe a material as constant per element or interpolated by a nodal-based Ansatz using shape functions defined on sub-blocks of the elements (e.g. \cite{Fichtner2009, Fichtner2013}). Nevertheless, this description is often closely tied to the spectral element method~(SEM) and not more than $p+1$ sub-blocks can be captured by the integration per element. The focus of the present contribution is to investigate the interactions between the wave field discretization and a completely independent material representation. Consequently, two key questions arise: Which high-order schemes are suitable to solve the forward and backward wave equation, and how can an independent yet efficient description of the material field be realized?
	
In general, high-order finite element methods outperform low-order approaches in approximating smooth wave solutions. Thanks to its diagonal mass matrix, the SEM is very popular in combination with explicit time stepping~\cite{Deville2002}. However, the diagonal structure of the mass matrix is a consequence of employing Lagrange basis functions in combination with Gauss Legendre Lobatto~(GLL) quadrature~\cite{Ronquist1987, Duczek2019}. If a material discretization is chosen that requires a different integration scheme, the diagonality and hence the central advantage of SEM in explicit time integration is lost. As an alternative to SEM, we use isogeometric analysis~(IGA)~\cite{Hughes2005} and an independent voxel-based material representation in the paper at hand to study the interaction between the wave field discretization and the resolution of the material field. The higher spline continuity across knot span boundaries allows to accurately solve wave problems with a significant lower number of degrees of freedom \cite{Hughes2014}. To easily incorporate complex geometric features of a structure, the finite cell method~(FCM)~\cite{Duester2017} is used. The isogeometric finite cell analysis (IGA-FCM) -- a combination of trimmed IGA and FCM -- has been previously studied in~\cite{Rank2012} and~\cite{Schillinger2012} in the context of linear elasticity, and has later been extended to dynamic problems~\cite{Leidinger2019,Messmer2022}. Similar approaches and in particular detailed mathematical analyses of IGA and immersed boundary methods are provided in the vast literature on CutFEM, e.g.~\cite{Burman2015, Burman2022}. 
	
The paper at hand combines an IGA-FCM approximation of the wave field with a voxelized representation of the material parameter. The subsequent key aspects of this multi-resolution FWI approach are addressed:
\begin{itemize}
	\item First, the forward simulation of the scalar wave equation using the consistent and row-sum lumped version of IGA-FCM is investigated -- with the goal to assess their suitability for an IGA-based FWI. 
	\item Second, the paper examines the interaction between the resolutions of the wave field and the material field for the inverse problem in terms of accuracy and computational cost. 
	\item Third, an adaptive locally refined material grid is introduced as part of the inversion process. The efficiency of this approach is demonstrated with 2D and 3D examples. 
\end{itemize}
This paper is the second contribution in a sequence applying immersed boundary methods to full waveform inversion. To ensure self-consistency, the basics introduced in the first paper~\cite{Buerchner2023} are briefly summarized. It is structured as follows: In~\secref{sec2}, we introduce the scalar wave equation, its spatial and temporal discretization, and the corresponding optimization problem. \secref{sec3} derives guidelines for the discretization of the wave field with IGA-FCM and its row-sum lumped variant for the forward wave problem. \secref{sec4} deals with the inverse problem, where the multi-resolution approach is evaluated on a 2D example. We then introduce an adaptive local refinement in the material field and apply the developed methodology to 2D and 3D examples. Finally, we conclude the paper in~\secref{sec5}. 
}

\FloatBarrier
\section{Full waveform inversion by isogeometric finite cell analysis}
{
\label{sec2}
\subsection{FCM for the scalar wave equation}
\FloatBarrier
	
We briefly summarize the nomenclature used in~\cite{Buerchner2023}, assuming an isotropic heterogeneous medium with density $\rho(\tensor{x})$ and wave speed $c(\tensor{x})$. Introducing the wave field $u(\tensor{x}, t)$, its acceleration $\ddot{u}(\tensor{x}, t)$ and the external force term $f(\tensor{x}, t)$, the scalar wave equation is defined on a computational domain $\Omega$ for the time $\mathcal{T} = [0, T_{\text{max}}]$ 	
\begin{equation}
	\rho(\tensor{x}) \ddot{u}(\tensor{x}, t) - \nabla \cdot \left( \rho(\tensor{x}) c^2(\tensor{x}) \nabla u(\tensor{x}, t) \right) = f(\tensor{x}, t)\text{,} \qquad \tensor{x} \in \Omega, t \in \mathcal{T} \text{.}
	\label{sweq}
\end{equation}
The initial conditions are $u(\tensor{x}, 0) = \dot{u}(\tensor{x}, 0) = 0$ for $\tensor{x} \in \Omega$, the boundary conditions $u(\tensor{x}, t) = 0\text{,} \, \tensor{x} \in \partial \Omega_\text{D}$ and $\tensor{n} \cdot \nabla u(\tensor{x},t) = 0\text{,} \, \tensor{x} \in \partial \Omega_\text{N}$ with $\partial \Omega = \partial \Omega_\text{D} \cup \partial \Omega_\text{N}$. We assume that a density $\rho_0$ and wave speed $c_0$ of the background material are given.
	
All known geometric features of a structure are incorporated in the initial domain $\Omega$, which may itself already have complex geometric shape. Applying the basic concept of the finite cell method, $\Omega$ is embedded in a larger, yet simply shaped domain $\Omega_e$. The original domain $\Omega$ is recovered through an indicator function $\alpha(\tensor{x})$, which assumes a small value $\epsilon$ (typically $10^{-5}$ to $10^{-8}$) representing a small density in the fictitious part of $\Omega_e$. While $\alpha(\tensor{x})$ is known a priori, unknown defects in the structure are iteratively identified by reconstructing a second, unknown scaling function $\gamma(\tensor{x})$, see \figref{embeddedDomain_withFlaw}. Since $\alpha$ and $\gamma$ only scale the density, the scalar wave equation takes the following form on the extended domain $\Omega_\text{e}$
\begin{equation}
		\alpha(\tensor{x}) \gamma(\tensor{x}) \rho_0 \ddot{u}(\tensor{x}, t) - \nabla \cdot \left( \alpha(\tensor{x}) \gamma(\tensor{x}) \rho_0 c_0^2 \nabla u(\tensor{x}, t) \right) = f(\tensor{x}, t)\text{,} \qquad \tensor{x} \in \Omega_e, t \in \mathcal{T} \text{.}
		\label{sweq_fcm}
\end{equation}
	
\begin{figure}[H]
	\vspace{-0.5cm}
	\centering
	\resizebox{0.8\textwidth}{!}{
		\input{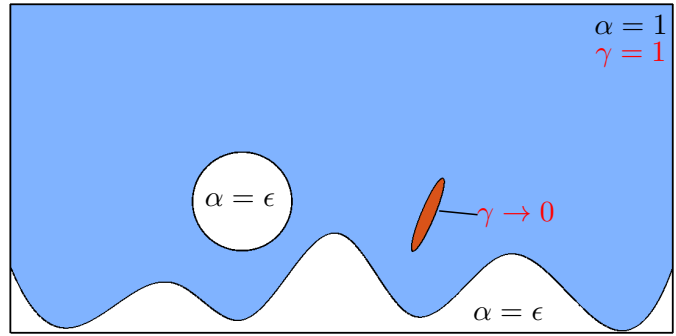}
	}
	\vspace{-1cm}
	\caption{A priori known geometry incorporated by the indicator function $\alpha$ with an unknown void reconstructed by the scaling~function~$\gamma$ (from~\cite{Buerchner2023})}
	\label{embeddedDomain_withFlaw}
\end{figure}


Note that this approach is independent of the basis chosen to discretize the spatial solution field. The spectral cell method~(SCM)~\cite{Duczek2014,Joulaian2014} uses Lagrange polynomials to approximate the wave field for both scalar and elastic wave equations. Isogeometric finite cell analysis (IGA-FCM) published in~\cite{Rank2012, Schillinger2012} can be combined with mass lumping to be used in explicit dynamics~\cite{Leidinger2020,Messmer2022,Stoter2023}. Other immersed boundary methods~(IBM) closely related to FCM are CutFEM~\cite{Burman2015, Sticko2020}, IBRA~\cite{Breitenberger2015}, aggregated FEM~\cite{Badia2018}, cgFEM~\cite{Nadal2013}, and the shifted boundary method~\cite{Main2018_1, Main2018_2}. Common to all these approaches is the idea to circumvent the task of boundary-conforming mesh generation by generating a non-boundary conforming computational grid and recovering the boundary at the level of the integration of the underlying bilinear forms. Obviously, this does not come at zero cost. In FCM, the integrands of the element mass and stiffness matrices are discontinuous for cells cut by boundaries of the physical domain. Several suitable integration approaches have been proposed to overcome this difficulty, including space-trees~\cite{Duester2008, Peto2020}, moment-fitting~\cite{Joulaian2016}, local integration meshes~\cite{Fries2016}, or smart octrees~\cite{Kudela2016}. In~\cite{Yang2012}, it is shown that FCM can be combined with a voxelized representation of the material parameter~$\alpha$. The integration is performed on a finer voxel grid using pre-integration to mitigate the computational burden of computing the system matrices~\cite{Korshunova2021}.

\FloatBarrier
\subsection{Spatial discretization of the wave field and material}
\FloatBarrier

For the approximation of the wave field $u(\tensor{x}, t)$, we use bivariate and trivariate B-spline discretizations in 2D and 3D~\cite{Hughes2005, Cottrell2009}. By defining a polynomial degree $p$ and a set of parametric coordinates, called the knot vector $\Xi = [\xi_1, \xi_2, ..., \xi_{n + p + 1}]$, the B-spline basis functions can be constructed -- where $\xi_i$ is the $i$th knot and $n$ the number of basis functions. Using the Cox-de Boor recursion formula~\cite{deBoor1978,Cohen2001}, the B-splines are
\begin{align}
	&N_{i, 0}(\xi) = \begin{cases}
	1, \quad \xi_i \leq \xi < \xi_{i+1} \\
	0, \quad \text{otherwise}
	\end{cases}, \qquad \text{\textbf{if} } p=0 \\
	&N_{i, p}(\xi) = \frac{\xi - \xi_i}{\xi_{i+p} - \xi_i} N_{i, p-1}(\xi) + \frac{\xi_{i+p+1} - \xi}{\xi_{i+p+1} - \xi_{i+1}} N_{i+1, p-1}(\xi) \qquad \text{\textbf{else}.}
\end{align}
The continuity $C^{p-k}$ of B-splines across the knot boundaries is defined by the knot multiplicity $k$. Henceforth, unless otherwise indicated we use open knot vectors with $k=1$ for all inner knots and $k=p+1$ for the end knot. With the set of all $n^\text{dof}$ bi- or trivariate basis functions $\tensor{N}$, the spatially discretized wave solution is 
\begin{equation}
	u(\tensor{x}, t) \approx \tilde{u}(\tensor{x}, t) = \sum_{i=1}^{n^\text{dof}} N_i(\tensor{x}) \hat{u}_i(t) = \tensor{N}(\tensor{x}) \hat{\tensor{u}}(t) \text{,}
\end{equation}
where $\hat{u}_i$ are the coefficients of the corresponding basis functions. Thanks to the non-negative partition of unity property of B-splines~\cite{Voet2022}, row-sum lumping is readily applicable and has been revived for boundary-conforming~\cite{Cottrell2006} and immersed IGA~\cite{Leidinger2020, Messmer2022}. Unfortunately, row-summing leads to a breakdown of $p$-convergence. As shown in~\cite{Cottrell2006}, the convergence of the first generalized eigenvalue is only of quadratic order for quadratic and cubic B-splines in 1D problems. Nevertheless, row-summing leads to a critical time step that becomes independent from the cut ratio of the knot spans if the physical domain is immersed~\cite{Leidinger2019, Messmer2022,Stoter2023} and, therefore -- at least at first sight -- seems to be an attractive option for explicit dynamics.

For the discretization of the material parameters, we utilize piecewise constant functions $N_{\text{m},i}$ defined on a voxel grid (see \figref{fig:discretizations}).
\begin{figure}[H]
	\centering
	\resizebox{0.6\textwidth}{!}{
		\input{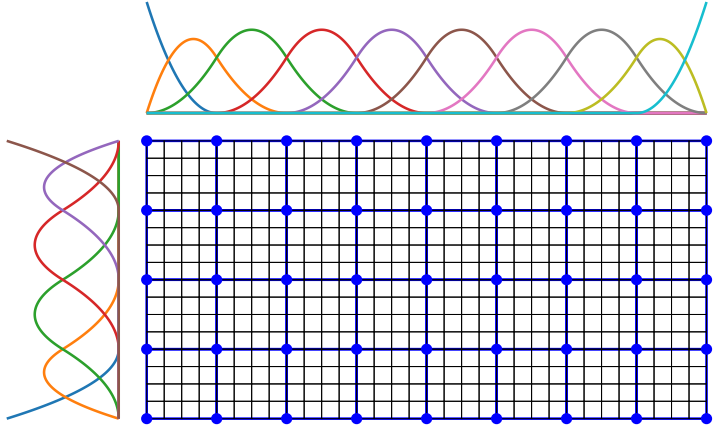}
	}
	\caption{Wave field mesh (thick blue lines and nodes) and material mesh (thin black lines)}
	\label{fig:discretizations}
\end{figure}
While this material grid can, in principle, be fully independent of the knot span grid for the spatial wave discretization, it is computationally advantageous to define it as a refinement with $n^\text{v}$ voxels per knot span in each spatial direction. With $n^\text{m}$ voxels discretizing the complete extended domain $\Omega_e$, the material discretization is 
\begin{equation}
	\gamma(\tensor{x}) \approx \tilde{\gamma}(\tensor{x}) = \sum_{i=1}^{n^\text{m}} N_{\text{m},i}(\tensor{x}) \hat{\gamma}_i = \tensor{N}_{\text{m}}(\tensor{x}) \hat{\tensor{\gamma}} \text{,}
\end{equation}
where $\hat{\gamma}_i$ denotes the value of the voxel. Since the material might be discontinuous within one knot span, the integration of the mass and stiffness matrices is performed by means of a composed integration at the voxel level, as e.g.~\cite{Yang2012}. 

\FloatBarrier
\subsection{Time integration}
\FloatBarrier

Derived by the Bubnov-Galerkin approach, we introduce the mass matrix $\tensor{M}$, the stiffness matrix $\tensor{K}$ and the external force vector $\tensor{F}$. The space-discrete form of the scalar wave equation is given by
\begin{equation}
	\tensor{M} \ddot{\hat{\tensor{u}}}(t) + \tensor{K}\hat{\tensor{u}}(t) = \hat{\tensor{f}}(t) \text{.}
\end{equation}
Applying second-order central differences~(CDM), the next time step $t_{i+1} = t_i + \Delta t$ is calculated from the previous two time steps $t_i$ and $t_{i-1} = t_i - \Delta t$:
\begin{equation}
	\hat{\tensor{u}}(t_{i+1}) = 2 \hat{\tensor{u}}(t_i) - \hat{\tensor{u}}(t_{i-1}) + \Delta t^2 \tensor{M}^{-1} \left[\hat{\tensor{f}}(t_i) - \tensor{K}\hat{\tensor{u}}(t_i)\right] \text{.}
\end{equation}
CDM is an explicit, conditionally stable time integration method. The number of time steps is denoted as $n^\text{t}$. The critical time step is given by 
\begin{equation}
    \Delta t_\text{c} = \frac{2}{\sqrt{\lambda_\text{max}(\tensor{K}, \tensor{M})}} \text{,}
\end{equation} 
where $\lambda_\text{max}(\tensor{K}, \tensor{M})$ is the largest eigenvalue of the generalized eigenproblem~\cite{Voet2022}. For details see e.g.~\cite{Hughes2012}.

\FloatBarrier
\subsection{Full waveform inversion}
\FloatBarrier

The goal of FWI is to find a set of unknown material coefficients $\hat{\tensor{\gamma}}$ to minimize the nonlinear optimization problem
\begin{equation}
	\hat{\tensor{\gamma}}^* = \arg \min_{\hat{\tensor{\gamma}}} \chi(\hat{\tensor{\gamma}}) \text{.}
\end{equation}
The cost function is defined by the squared residual between simulation and experiments, summed up over $n^\text{r}$ receiver positions in $n^\text{s}$ experiments
\begin{equation}
	\chi(\hat{\tensor{\gamma}}) = \frac{1}{2} \sum_{s=1}^{n^\text{s}} \sum_{r=1}^{n^\text{r}} \int_T \int_\Omega \left[ \left( u^s(\hat{\tensor{\gamma}}; \tensor{x}, t) - u^{0, s}(\tensor{x}, t) \right)^2 \delta(\tensor{x} - \tensor{x}^r) \right] d\Omega dt \text{,}
\end{equation}
where $u^s(\hat{\tensor{\gamma}}; \tensor{x}, t)$ is the solution of a wave simulation with the current material $\hat{\tensor{\gamma}}$ and $u^{0, s}(\tensor{x}^r, t)$ is the corresponding experimental measurement at the receiver position $\tensor{x}^r$. A typical experimental setup can be found in~\cite{Buerchner2023}, and the computation of the gradient applying the adjoint method is derived according to~\cite{Fichtner2006a, Fichtner2006b}. In the following, we revise the derived formulas from \cite{Buerchner2023}. The sensitivity kernel with respect to the scaling function $\gamma$ for a given set $\hat{\tensor{\gamma}}$ is
\begin{equation}
	K_\gamma(\tensor{x}) = \sum_{s=1}^{n^\text{s}} \int_T \left[ - \alpha(\tensor{x}) \rho_0 \dot{u}^{s,\dagger}(\hat{\tensor{\gamma}}; \tensor{x}, t) \dot{u}^s(\hat{\tensor{\gamma}}; \tensor{x}, t) + \alpha(\tensor{x}) \rho_0 c_0^2 \nabla u^{s,\dagger}(\hat{\tensor{\gamma}}; \tensor{x}, t) \cdot \nabla u^s(\hat{\tensor{\gamma}}; \tensor{x}, t) \right] dt \text{,}
\end{equation}
where $u^{s,\dagger}$ is the adjoint solution of experiment $s$. The gradient with respect to the voxel coefficients $\gamma_i$ is approximated by evaluating the sensitivity kernel at the voxel mid positions $\tensor{x}_{\hat{\gamma},i}$
\begin{equation}
	\frac{d \chi}{d \hat{\gamma}_i} \approx \int_\Omega K_\gamma \delta (\tensor{x}_{\hat{\gamma},i} - \tensor{x}) d \Omega = K_\gamma(\tensor{x}_{\hat{\gamma},i})
\end{equation}
or in discretized form
\begin{equation}
		\frac{d \chi}{d \hat{\gamma}_i} \approx \sum_{s=1}^{n^\text{s}} \int_\Omega \int_T \left[ - \rho_0 (\dot{\hat{\tensor{u}}}^{s, \dagger})^T \tensor{N}^T \tensor{N} \dot{\hat{\tensor{u}}}^s + \rho_0 c_0^2 (\hat{\tensor{u}}^{s,\dagger})^T \tensor{B}^T \tensor{B} \hat{\tensor{u}}^s \right] dt \delta (\tensor{x}_{\hat{\gamma},i} - \tensor{x}) d \Omega \label{eq:sensitivity}
\end{equation}
The unknown material field $\gamma$ is optimized only within the physical domain. Therefore, it is assumed that the indicator function $\alpha$ at the considered positions $\tensor{x}_{\hat{\gamma},i}$ is equal to $1$ and, consequently that it vanishes in the above equation. In gradient-based optimization, the material is iteratively improved with an update step $\Delta \hat{\tensor{\gamma}}$. The superscript $k$ denotes the current iteration
\begin{equation}
	\hat{\tensor{\gamma}}^{(k+1)} = \hat{\tensor{\gamma}}^{(k)} + \Delta \hat{\tensor{\gamma}}^{(k)}
\end{equation}
Quasi-Newton type methods take into account the current gradient $\nabla_{\hat{\tensor{\gamma}}} \chi(\hat{\tensor{\gamma}}^{(k)})$ and an approximate of the inverse Hessian $\tensor{H}_a^{-1}(\hat{\tensor{\gamma}}^{k})$ in the model update 
\begin{equation}
	\Delta \hat{\tensor{\gamma}}^{(k)} = -\tensor{H}_a^{-1}(\hat{\tensor{\gamma}}^{k}) \nabla_{\hat{\tensor{\gamma}}} \chi(\hat{\tensor{\gamma}}^{(k)}) \text{.}
\end{equation}
For an introduction to gradient-based optimization, we refer to~\cite{Nocedal2006}. In the paper at hand, the matrix-free and bounded L-BFGS-B of the Python library SciPy~\cite{2020SciPy} is applied. 

The computational cost and memory requirement of the gradient computation can be readily estimated. Considering~\eqref{eq:sensitivity}, the sensitivity kernel must be computed for all $n^\text{m}$ voxel mid points. Assuming that the forward and adjoint solutions $\hat{\tensor{u}}^s$ and $\hat{\tensor{u}}^{s,\dagger}$ have been computed and are temporarily stored for all time steps $n^\text{t}$, the integrand is evaluated and summed up for all time steps. Since the evaluation is done locally for every voxel, only the $n^\text{dof,local}$ coefficients associated with the basis functions being non-zero in the corresponding knot span need to be considered. For the chosen wave field discretization (see~\figref{fig:discretizations}), $(p+1)^d$ non-zero basis functions exist at every position in space, where $d$ is the spatial dimension and $p$ is the chosen polynomial order.
To summarize, the computational effort is 
\begin{equation}
    \mathcal{O}(n^\text{s} \times n^\text{t} \times n^\text{dof,local} \times n^\text{m}) = \mathcal{O}(n^\text{s} \times n^\text{t} \times p^d \times n^\text{m}) \text{.}
    \label{eq:O_time}
\end{equation}
Memory requirement is 
\begin{equation}
    \mathcal{O}(n^\text{t} \times n^\text{dof}) \text{.}
    \label{eq:O_memory}
\end{equation}

For a chosen polynomial degree~$p$, the computational effort depends only on the \textit{total} number of voxels~$n^\text{m}$, and on the relation between the size of the knot span and the voxel size. Since the computation of the gradient consumes a large fraction of the overall effort, insight into an optimized relationship between the resolutions of the wave field and the material grid is desirable~(see \secref{sec4}).
}

\FloatBarrier
\section{Solving the wave equation by IGA-FCM}
{
\label{sec3}
\FloatBarrier
In this section, we first investigate whether IGA-FCM is generally suitable as a wave equation solver in the framework of full waveform inversion. The following observations are important:
\begin{enumerate}
    \item A wave equation solver should allow a high convergence rate which can even be selected depending on the expected smoothness of the wave field. It is well known \cite{Hughes2005,Hughes2014} that increasing the polynomial degree of the Ansatz functions outperforms a refinement of the mesh (h-extension) by far. Therefore, IGA (like other high-order solvers) is well qualified in the context of consideration. This applies in particular in combination with immersed methods such as the FCM, as restrictions on the geometric shape of the domain $\Omega$ are minimal. 
    \item A solver should provide high accuracy per degree of freedom. $k$-extension of IGA (see \cite{Hughes2005}) combines the increase of the polynomial degree and the increase of the smoothness of the Ansatz in such a way that one degree of freedom per spatial direction is enough to gain one additional order of convergence. The computational effort associated with this additional degree of freedom may be significant due to a loss of sparsity and an increase of fill-in throughout the system matrices. Yet, the additional computational cost is offset by a drastic reduction in memory requirements. This is important because, in the adjoint gradient computation, the coefficients of the solution vectors of all time steps of the forward simulation must be stored temporarily.
    \item The goal of FWI is to identify geometric features that may be small compared to the mesh size of the wave field discretization. This can be achieved (as will be shown in the following section) by using a material grid that is refined compared to the mesh of the wave field. 
\end{enumerate}

Let us now take a look at the IGA-FCM solution of the scalar wave problem using consistent and lumped mass matrices. We consider a two-dimensional domain of $l_x = 10$ and $l_y = 5$ with a circular hole of radius $r = 0.5$ at position $x_c = 6$ and $y_c = 2.85$, shown in \figref{fig:forwardExampleSetting}. The density and wave speed of the background material are set to $\rho_0 = 1$ and $c_0 = 1$. A 2-cycle sine burst 
\begin{equation}
    g(t) = \begin{cases}
        \sin{(2 \pi f t)} \sin{\left(\frac{\pi f}{2}\right)} &\qquad \text{, } t \leq \frac{2}{f} \\
        0 &\qquad \text{, else}
    \end{cases}
\end{equation}
with a central frequency $f=0.5$ and a spatial Gaussian distribution 
\begin{equation}
    f(x,y) = e^{-\left( \frac{(x-x_s)^2}{2\sigma_x^2} + \frac{(y-y_s)^2}{2\sigma_y^2} \right)}
\end{equation}
is excited at position $x_s = 2$ and $y_s = 2.5$ with $\sigma_x = \sigma_y = 0.25$, leading to a dominant wavelength $\lambda_\text{dom} = 2$. The wave propagation is computed for $T_\text{max} = 10$.
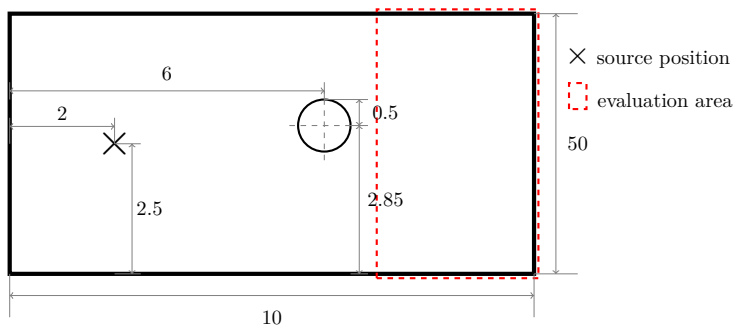
\begin{figure}[H]
	\centering
	\resizebox{0.6\textwidth}{!}{
		\begin{tikzpicture}
	\centering
	\filldraw[draw=black,fill=white, line width=0.8mm] (0.0,0.0) rectangle (0.6\textwidth,0.3\textwidth);
	\filldraw[draw=red,fill=none, line width=0.4mm, dashed] (0.42\textwidth,-0.005\textwidth) rectangle (0.605\textwidth,0.305\textwidth);
	\draw [draw=gray, line width=0.1mm] (0.6\textwidth,0.0) -- (0.65\textwidth, 0.0);
	
	\draw [draw=gray, line width=0.1mm] (0.6\textwidth,0.3\textwidth) -- (0.65\textwidth,
	0.3\textwidth);
	\draw[draw=gray, <->] (0.625\textwidth,0.0) -- (0.625\textwidth,0.3\textwidth);
	\node at (0.65\textwidth,0.15\textwidth) {$50$};
	\draw [draw=gray, line width=0.1mm] (0.0,0.0) -- (0.0, -0.05\textwidth);
	\draw [draw=gray, line width=0.1mm] (0.6\textwidth,0.0) -- (0.6\textwidth, -0.05\textwidth);
	\draw[draw=gray, <->] (0.0,-0.025\textwidth) -- (0.6\textwidth,-0.025\textwidth);
	\node at (0.3\textwidth,-0.05\textwidth) {$10$};
	
	\filldraw[draw=black,fill=white, line width=0.4mm] (0.36\textwidth,0.171\textwidth) circle (0.03\textwidth);
	
	\draw [draw=gray, dashed, line width=0.2mm] (0.36\textwidth,0.131\textwidth) -- (0.36\textwidth, 0.211\textwidth);
	\draw [draw=gray, line width=0.1mm] (0.36\textwidth,0.201\textwidth) -- (0.36\textwidth, 0.221\textwidth);
	\draw [draw=gray, line width=0.1mm] (0.0\textwidth,0.201\textwidth) -- (0.0\textwidth, 0.221\textwidth);
	\draw[draw=gray, <->] (0.0\textwidth,0.211\textwidth) -- (0.36\textwidth, 0.211\textwidth);
	\node at (0.18\textwidth,0.231\textwidth) {$6$};
	
	\draw [draw=gray, dashed, line width=0.2mm] (0.32\textwidth,0.171\textwidth) -- (0.40\textwidth, 0.171\textwidth);
	\draw [draw=gray, line width=0.1mm] (0.39\textwidth,0.0) -- (0.41\textwidth, 0.0);
	\draw [draw=gray, line width=0.1mm] (0.39\textwidth,0.171\textwidth) -- (0.41\textwidth, 0.171\textwidth);
	\draw[draw=gray, <->] (0.4\textwidth,0.0) -- (0.4\textwidth,0.171\textwidth);
	\node at (0.43\textwidth,0.0855\textwidth) {$2.85$};
	
	\draw [draw=gray, line width=0.1mm] (0.36\textwidth,0.201\textwidth) -- (0.41\textwidth, 0.201\textwidth);
	\draw[draw=gray, <->] (0.4\textwidth,0.201\textwidth) -- (0.4\textwidth,0.171\textwidth);
	\node at (0.43\textwidth,0.186\textwidth) {$0.5$};
	
	\node at (0.12\textwidth,0.15\textwidth) {\Huge$\times$};
	\node at (0.73\textwidth,0.25\textwidth) {{\LARGE$\times$} source position};
	\filldraw[draw=red,fill=none, line width=0.4mm, dashed] (0.64\textwidth,0.19\textwidth) rectangle (0.66\textwidth,0.22\textwidth);
	\node at (0.75\textwidth,0.2\textwidth) {evaluation area};
	
	\draw [draw=gray, line width=0.1mm] (0.12\textwidth,0.18\textwidth) -- (0.12\textwidth, 0.15\textwidth);
	\draw [draw=gray, line width=0.1mm] (0.0\textwidth,0.18\textwidth) -- (0.0\textwidth, 0.15\textwidth);
	\draw[draw=gray, <->] (0.0\textwidth,0.17\textwidth) -- (0.12\textwidth, 0.17\textwidth);
	\node at (0.06\textwidth,0.185\textwidth) {$2$};
	
	\draw [draw=gray, line width=0.1mm] (0.12\textwidth,0.15\textwidth) -- (0.15\textwidth, 0.15\textwidth);
	\draw [draw=gray, line width=0.1mm] (0.12\textwidth,0.0\textwidth) -- (0.15\textwidth, 0.0\textwidth);
	\draw[draw=gray, <->] (0.14\textwidth,0.0) -- (0.14\textwidth,0.15\textwidth);
	\node at (0.16\textwidth,0.075\textwidth) {$2.5$};
\end{tikzpicture}
	}
	\caption{2D domain with a hole}
	\label{fig:forwardExampleSetting}
\end{figure}
	
To compare the accuracy of consistent and lumped IGA-FCM for different polynomial orders, the wave solution at $T_{max}$ is evaluated in the marked area $\left[ 7, 10 \right] \times \left[ 0, 5 \right]$ on the right side of the hole. For this purpose, this area is sampled with $N_\text{e} = 601 \times 1001$ equidistant evaluation points in $x$- and $y$-direction. These evaluation points correspond to an arbitrary number of receiver positions in the FWI. The normalized error with respect to an overkill reference solution $u_\text{ref}$
\begin{equation}
    \epsilon = \frac{\sqrt{\sum_{e=0}^{N_\text{e}} \left( u(\tensor{x}_e, T_\text{max}) - u_\text{ref}(\tensor{x}_e, T_\text{max}) \right)^2 }}{\sqrt{\sum_{e=0}^{N_\text{e}} \left( u_\text{ref}(\tensor{x}_e, T_\text{max}) \right)^2 }} \text{.}
\end{equation}
\FloatBarrier%
can be interpreted as an error proportional to the $\mathcal{L}_2$ error calculated using the Riemann sum for integration. The reference solution is obtained with quintic $C^0$ continuous integrated Legendre polynomials defined on a mesh of mesh size $h = \frac{1}{16}$, resulting in $160$ elements in $x$-direction and $80$ elements in $y$-direction. The FCM indicator function defining the physical part of the computational domain is set to $\alpha = 10^{-8}$ inside the hole. For the consistent and lumped version of IGA-FCM, the solution is computed for linear, quadratic, cubic and quartic $C^{p-1}$ B-splines. The mesh size is varied using $h = \frac{1}{2}, \frac{1}{4}, \frac{1}{8}, \frac{1}{16}, \frac{1}{32}$. In order to reduce spatial and temporal integration errors to a minimum, all simulations are carried out with $n^\text{t} = 100\,000$ time steps -- and the integration of the mass and stiffness matrices and the force vector is performed by a quadtree-quadrature applying a depth of $d = 10$. \figref{fig:forwardExampleConvergence} shows the results for the consistent version of IGA-FCM, referred to as `c-IGA-FCM', and for the lumped version of IGA-FCM, denoted as `l-IGA-FCM'. Reference lines proportional to $h^2$, $h^3$, $h^4$, and $h^5$ are depicted. 

\begin{figure}[H]
	\centering
	\resizebox{0.8\textwidth}{!}{
		\begin{tikzpicture}
	\begin{loglogaxis}[
		xmin = 1/40, xmax = 0.55,
		ymin = 2e-8, ymax = 2,
		xtick={0.05, 0.1, 0.2, 0.4},
		xticklabels = {$0.05$, $0.1$, $0.2$, $0.4$},
		ytick={0.000001, 0.0001, 0.01, 1},
		grid=both,
		width = 1.0\textwidth,
		height = 0.75\textwidth,
		xlabel = {mesh size $h$},
		ylabel style={align=center}, ylabel=error~$\epsilon$,
		legend style={at={(1,0)}, anchor=south east}]
		
		\addplot[black, mark = x, line width=0.5pt] file[] {tikzpictures/data/cIGA_p1.dat};
		\addplot[black, mark = *, line width=0.5pt] file[] {tikzpictures/data/cIGA_p2.dat};
		\addplot[black, mark = square, line width=0.5pt] file[] {tikzpictures/data/cIGA_p3.dat};
		\addplot[black, mark = o, line width=0.5pt] file[] {tikzpictures/data/cIGA_p4.dat};
		
		\addplot[blue, mark = x, line width=0.5pt] file[] {tikzpictures/data/lIGA_p1.dat};
		\addplot[blue, mark = *, line width=0.5pt] file[] {tikzpictures/data/lIGA_p2.dat};
		\addplot[blue, mark = square, line width=0.5pt] file[] {tikzpictures/data/lIGA_p3.dat};
		\addplot[blue, mark = o, line width=0.5pt] file[] {tikzpictures/data/lIGA_p4.dat};
		
		\addplot[green, line width=0.5pt] file[] {tikzpictures/data/ref_p1.dat};
		\addplot[green, line width=0.5pt] file[] {tikzpictures/data/ref_p2.dat};
		\addplot[green, line width=0.5pt] file[] {tikzpictures/data/ref_p3.dat};
		\addplot[green, line width=0.5pt] file[] {tikzpictures/data/ref_p4.dat};
		
		\legend{c-IGA-FCM $p=1$,c-IGA-FCM $p=2$, c-IGA-FCM $p=3$, c-IGA-FCM $p=4$, l-IGA-FCM $p=1$,l-IGA-FCM $p=2$, l-IGA-FCM $p=3$, l-IGA-FCM $p=4$},
	\end{loglogaxis}
\node[] at (0.175\textwidth,0.44\textwidth) {$\propto h^2$};
\node[] at (0.175\textwidth,0.2\textwidth) {$\propto h^3$};
\node[] at (0.175\textwidth,0.07\textwidth) {$\propto h^4$};
\node[] at (0.4\textwidth,0.07\textwidth) {$\propto h^5$};
\end{tikzpicture}
	}
	\caption{Convergence for consistent and lumped IGA-FCM with $p=1, 2, 3, 4$}
	\label{fig:forwardExampleConvergence}
\end{figure}
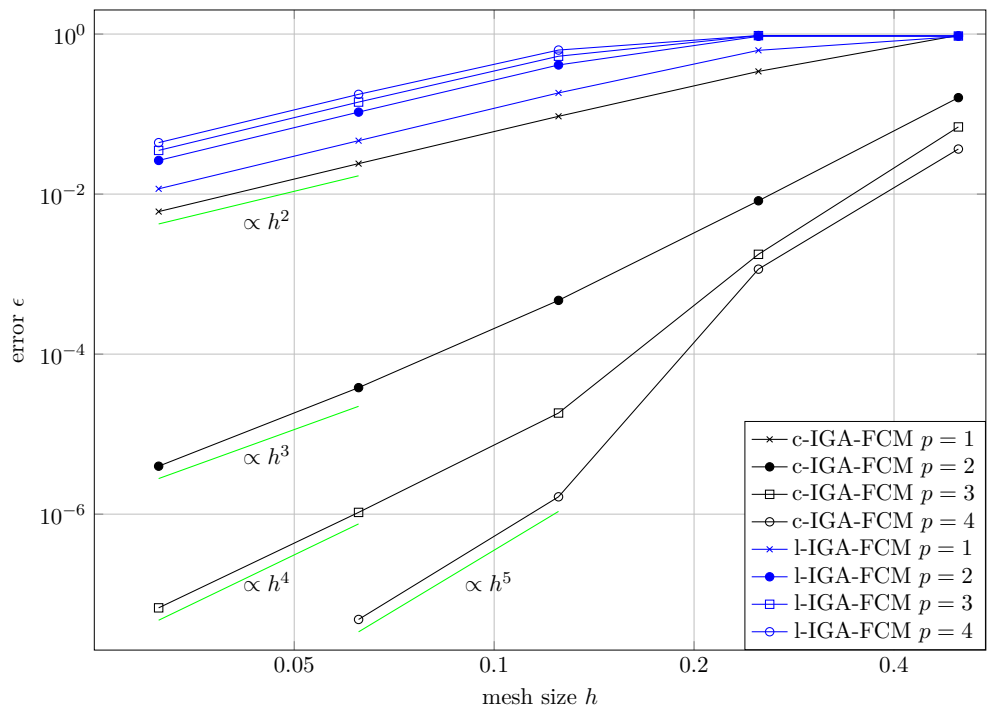

Corresponding to the results concerning the generalized eigenvalue problem in \cite{Cottrell2006}, a collapse of p-convergence can be observed for the problem at hand if the lumped version of IGA-FCM is applied. The order of convergence remains quadratic regardless of the polynomial order $p$, while the convergence constant even deteriorates as $p$ increases. For the consistent version of IGA-FCM, however, we observe an increase in the order of convergence. As expected from results concerning the eigenfunction of the generalized eigenvalue problem, the error in time has an asymptotic convergence of $\mathcal{O}(h^{p+1})$ as well. Additionally, improvement in the constant can be noticed as $p$ increases. Since FWI requires to temporarily store full wave fields, a memory-efficient solver is of great advantage. Thus, lumped IGA-FCM is not suitable for this application -- and we will not consider it for the inverse problem, due to the higher number of degrees of freedom that is required to achieve a desired accuracy. However, consistent IGA-FCM provides a very memory-efficient and accurate solution of the wave problem and, therefore, is our method of choice from here on.

\FloatBarrier
}

\section{The inverse problem}
{
\label{sec4}
\FloatBarrier

\subsection{Multi-resolution approach}
\FloatBarrier

To evaluate the applicability of the multi-resolution approach, we consider the embedded domain example of \figref{embeddedDomain_withFlaw}. As given in \cite{Buerchner2023}, the sample has a size of $\SI{100}{\milli \meter} \times \SI{50}{\milli \meter}$, density and wave speed are $\SI{2700}{\kilogram \per \meter^3}$ and $\SI{6000}{\meter \per \second}$, the lower boundary of the physical domain is defined by cubic splines interpolating the nine points $(0, \SI{10}{\milli \meter})$, $(\SI{10}{\milli \meter}, \SI{1}{\milli \meter})$, $(\SI{25}{\milli \meter}, \SI{7.5}{\milli \meter})$, $(\SI{35}{\milli \meter}, \SI{2}{\milli \meter})$, $(\SI{50}{\milli \meter}, \SI{15}{\milli \meter})$, $(\SI{60}{\milli \meter}, \SI{3}{\milli \meter})$, $(\SI{75}{\milli \meter}, \SI{12}{\milli \meter})$, $(\SI{90}{\milli \meter}, \SI{1}{\milli \meter})$, and $(\SI{100}{\milli \meter}, \SI{10}{\milli \meter})$, the circular hole is centered at $(\SI{35}{\milli \meter}, \SI{20}{\milli \meter})$ with radius $r=\SI{7.5}{\milli \meter}$, and the unknown ellipse is located at $(\SI{63}{\milli \meter}, \SI{18}{\milli \meter})$ with semi-axes $a=\SI{6}{\milli \meter}$ and $b=\SI{1}{\milli \meter}$, rotated by $\SI{67.5}{\degree}$. The sample is excited by $17$ sources centered at the top surface with a spacing of $\SI{4}{\milli \meter}$. Whenever a signal is sent from one of the sources, all source locations are used as receiver positions, mimicking the functionality of physical transducers. The central frequency of the 2-cycle sine burst is $f=\SI{500}{\kilo \hertz}$, corresponding to a dominant wavelength $\lambda_\text{dom} = \SI{12}{\milli \meter}$. The synthetic reference data are computed with a boundary-conforming mesh of linear quadrilateral elements with over $50 \frac{\text{dof}}{\lambda_\text{dom}}$. The inversion is done using full matrix capture (FMC, see \cite{Clark2004}) including signals of all sources, and a maximum of $10$ iterations is performed. No regularization of the inverse problem beyond the intrinsic one associated to the discretization with B-splines is applied.

 Taking into account the results of \secref{sec3}, the polynomial degree of the wave field is chosen to be $p = 2$ and $p = 3$. Wave field and material grids are discretized independently. The knot span length of the wave field mesh is varied between $h = 5 \si{\milli \meter}$, $2.5 \si{\milli \meter}$, and $1.25 \si{\milli \meter}$, leading to discretizations with $2.4$, $4.8$, and $9.6$ knot spans per wavelength. These meshes are combined with independent grids of voxel size with $h^\text{v} = 1.25\si{\milli \meter}$, $0.625\si{\milli \meter}$, or $0.3125\si{\milli \meter}$. The nine resulting combinations of the wave field and material grids are listed in \tabref{Tab:tab1}. The number of voxels in each dimension per knot span is denoted as $n^\text{v}$. In order to incorporate the a priori known geometry the integration of the system matrices is carried out using a quadtree of depth $d = p + 1$ on each cut knot span. Inside the void domain, $\alpha$ is set to $10^{-5}$ and the inversion of $\gamma$ is bounded between $\gamma_\text{min} = 10^{-5}$ and $\gamma_\text{max} = 1$. The wave field is computed for a time span of $\SI{6.0 e-5}{\second}$ in $3000$ time steps.
 
\begin{table}[H]
\centering
  \caption{Wave field and material grid sizes}
  \renewcommand{\arraystretch}{1.5}
  \begin{tabular}{|l|c|c|c|}
  \hline
  & $h^\text{v} = 1.25 \si{\milli \meter}$ & $h^\text{v} = 0.625 \si{\milli \meter}$ & $h^\text{v} = 0.3125 \si{\milli \meter}$ \\
  \hline
  \hline$h = 5 \si{\milli \meter}$ & $n^\text{v} = 4$ & $n^\text{v} = 8$ & $n^\text{v} = 16$ \\
  \hline
  $h = 2.5 \si{\milli \meter}$ & $n^\text{v} = 2$ & $n^\text{v} = 4$ & $n^\text{v} = 8$ \\
  \hline
  $h = 1.25 \si{\milli \meter}$ & $n^\text{v} = 1$ & $n^\text{v} = 2$ & $n^\text{v} = 4$ \\
  \hline
  \end{tabular}
  \label{Tab:tab1}
\end{table}

\figref{noRefinementInversions_p2} shows the inversion results for $p=2$, \figref{noRefinementInversions_p3} for $p=3$, and \tabref{Tab:tab_p2} and \tabref{Tab:tab_p3} list the computation times. For the graphical representation, the material field is visualized throughout the computational domain on the level of the applied voxel size. As noted above, the a priori known geometric features (i.e., the lower boundary and the circular hole) are resolved more precisely in the FCM computation by a quadtree integration. 
\FloatBarrier%

\begin{figure}
	\centering
	\begin{subfigure}[t]{0.3\textwidth}
		\centering
		\includegraphics[width=\textwidth]{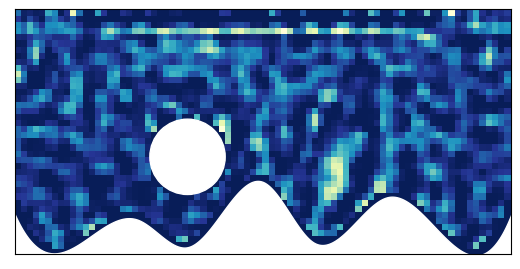}
		\caption{$h = 5 \si{\milli \meter}$, $n^\text{v} = 4$}
		\label{fig:noRefinementInversions_p2_1}
	\end{subfigure}
	\begin{subfigure}[t]{0.3\textwidth}
		\centering
		\includegraphics[width=\textwidth]{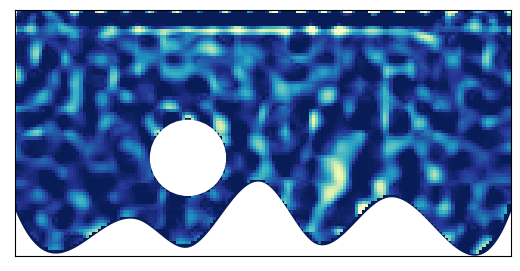}
		\caption{$h = 5 \si{\milli \meter}$, $n^\text{v} = 8$}
	\end{subfigure}
	\begin{subfigure}[t]{0.3\textwidth}
		\centering
		\includegraphics[width=\textwidth]{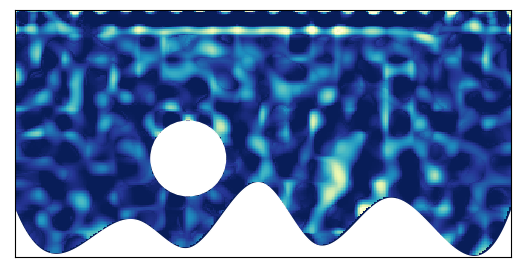}
		\caption{$h = 5 \si{\milli \meter}$, $n^\text{v} = 16$}
		\label{fig:noRefinementInversions_p2_3}
	\end{subfigure}
	\begin{subfigure}[t]{0.3\textwidth}
		\centering
		\includegraphics[width=\textwidth]{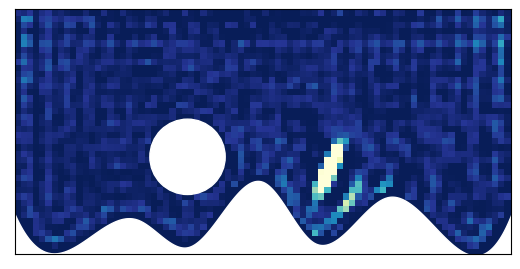}
		\caption{$h = 2.5 \si{\milli \meter}$, $n^\text{v} = 2$}
		\label{fig:noRefinementInversions_p2_4}
	\end{subfigure}
	\begin{subfigure}[t]{0.3\textwidth}
		\centering
		\includegraphics[width=\textwidth]{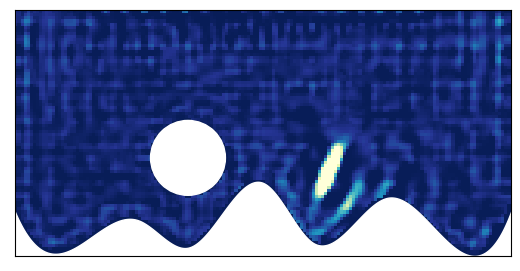}
		\caption{$h = 2.5 \si{\milli \meter}$, $n^\text{v} = 4$}
	\end{subfigure}
	\begin{subfigure}[t]{0.3\textwidth}
		\centering
		\includegraphics[width=\textwidth]{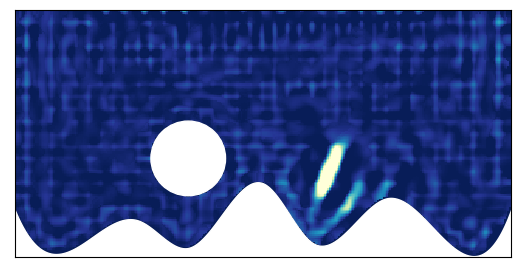}
		\caption{$h = 2.5 \si{\milli \meter}$, $n^\text{v} = 8$}
		\label{fig:noRefinementInversions_p2_6}
	\end{subfigure}
	\begin{subfigure}[t]{0.3\textwidth}
		\centering
		\includegraphics[width=\textwidth]{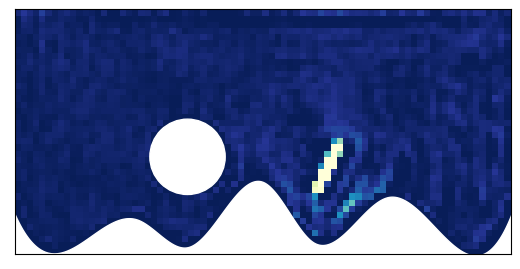}
		\caption{$h = 1.25 \si{\milli \meter}$, $n^\text{v} = 1$}
		\label{fig:noRefinementInversions_p2_7}
	\end{subfigure}
	\begin{subfigure}[t]{0.3\textwidth}
		\centering
		\includegraphics[width=\textwidth]{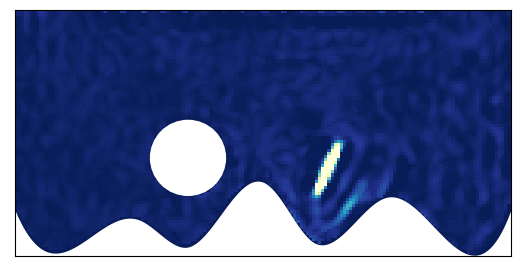}
		\caption{$h = 1.25 \si{\milli \meter}$, $n^\text{v} = 2$}
	\end{subfigure}
	\begin{subfigure}[t]{0.3\textwidth}
		\centering
		\includegraphics[width=\textwidth]{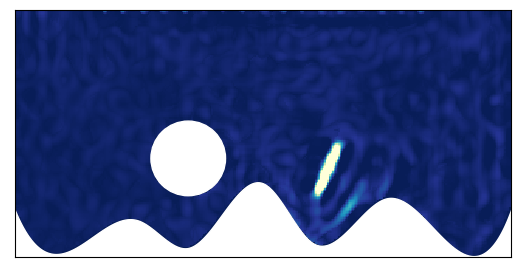}
		\caption{$h = 1.25 \si{\milli \meter}$, $n^\text{v} = 4$}
		\label{fig:noRefinementInversions_p2_9}
	\end{subfigure}
	\caption{Inversion results for polynomial degree $p=2$}
	\label{noRefinementInversions_p2}
\end{figure}

\begin{figure}
	\centering
	\begin{subfigure}[t]{0.3\textwidth}
		\centering
		\includegraphics[width=\textwidth]{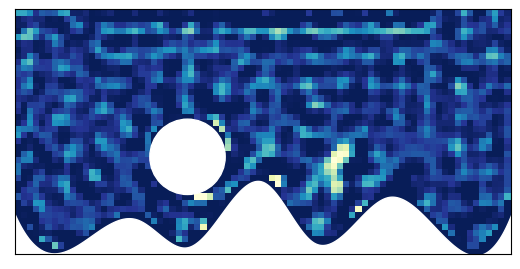}
		\caption{$h = 5 \si{\milli \meter}$, $n^\text{v} = 4$}
		\label{fig:noRefinementInversions_p3_1}
	\end{subfigure}
	\begin{subfigure}[t]{0.3\textwidth}
		\centering
		\includegraphics[width=\textwidth]{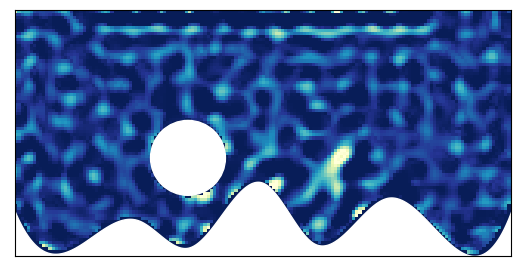}
		\caption{$h = 5 \si{\milli \meter}$, $n^\text{v} = 8$}
	\end{subfigure}
	\begin{subfigure}[t]{0.3\textwidth}
		\centering
		\includegraphics[width=\textwidth]{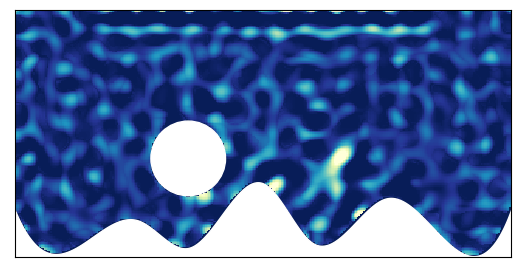}
		\caption{$h = 5 \si{\milli \meter}$, $n^\text{v} = 16$}
		\label{fig:noRefinementInversions_p3_3}
	\end{subfigure}
	\begin{subfigure}[t]{0.3\textwidth}
		\centering
		\includegraphics[width=\textwidth]{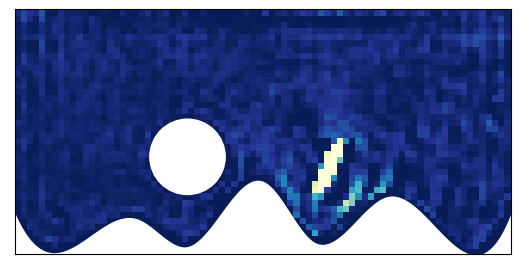}
		\caption{$h = 2.5 \si{\milli \meter}$, $n^\text{v} = 2$}
		\label{fig:noRefinementInversions_p3_4}
	\end{subfigure}
	\begin{subfigure}[t]{0.3\textwidth}
		\centering
		\includegraphics[width=\textwidth]{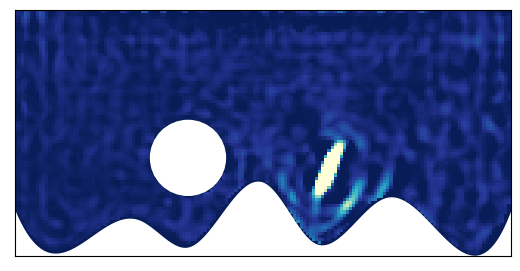}
		\caption{$h = 2.5 \si{\milli \meter}$, $n^\text{v} = 4$}
	\end{subfigure}
	\begin{subfigure}[t]{0.3\textwidth}
		\centering
		\includegraphics[width=\textwidth]{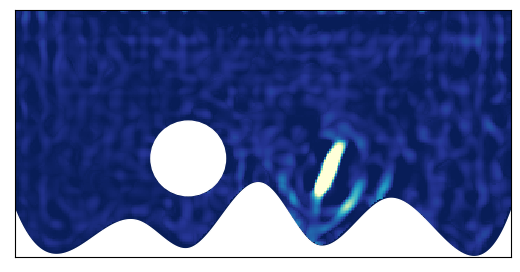}
		\caption{$h = 2.5 \si{\milli \meter}$, $n^\text{v} = 8$}
		\label{fig:noRefinementInversions_p3_6}
	\end{subfigure}
	\begin{subfigure}[t]{0.3\textwidth}
		\centering
		\includegraphics[width=\textwidth]{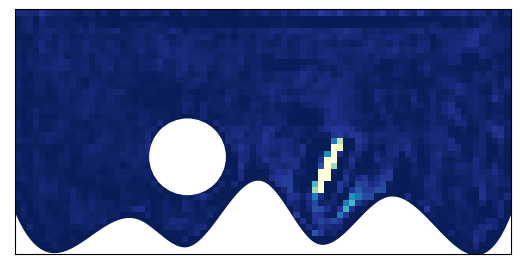}
		\caption{$h = 1.25 \si{\milli \meter}$, $n^\text{v} = 1$}
		\label{fig:noRefinementInversions_p3_7}
	\end{subfigure}
	\begin{subfigure}[t]{0.3\textwidth}
		\centering
		\includegraphics[width=\textwidth]{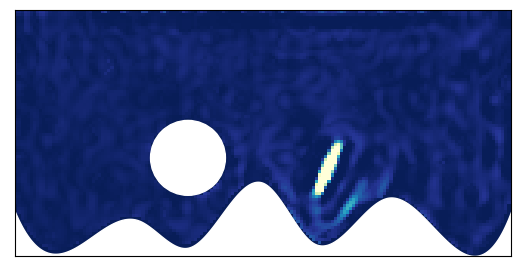}
		\caption{$h = 1.25 \si{\milli \meter}$, $n^\text{v} = 2$}
	\end{subfigure}
	\begin{subfigure}[t]{0.3\textwidth}
		\centering
		\includegraphics[width=\textwidth]{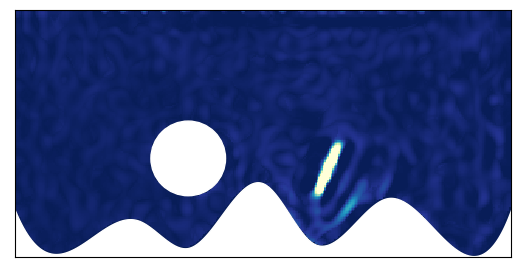}
		\caption{$h = 1.25 \si{\milli \meter}$, $n^\text{v} = 4$}
		\label{fig:noRefinementInversions_p3_9}
	\end{subfigure}
	\caption{Inversion results for polynomial degree $p=3$}
	\label{noRefinementInversions_p3}
\end{figure}

If the wave field discretization is too coarse, dispersion errors cause imprecise wave simulations and consequently large artifacts in the material reconstruction. Regardless of the material resolution, the defect cannot be identified with a knot span size of $h = 5 \si{\milli\meter}$ and $p = 2$, see \figsref{fig:noRefinementInversions_p2_1}~to~\ref{fig:noRefinementInversions_p2_3}. The same holds for $p = 3$, see \figsref{fig:noRefinementInversions_p3_1}~to~\ref{fig:noRefinementInversions_p3_3}. Increasing the total number of knot spans improves the quality of the wave simulations and, consequently, mitigates the artifacts occurring in the inversions. While $h = 2.5 \si{\milli\meter}$ and $p = 2$ (\figsref{fig:noRefinementInversions_p2_4}~to~\ref{fig:noRefinementInversions_p2_6}) still lead to clearly visible artifacts, in particular to the right of the defect, $p = 3$ leads to a defect that is well distinguishable from the background (\figsref{fig:noRefinementInversions_p3_4}~to~\ref{fig:noRefinementInversions_p3_6}). For the finest meshes, the background perturbations almost completely disappear, see \figsref{fig:noRefinementInversions_p2_7}~to~\ref{fig:noRefinementInversions_p2_9} for $p = 2$ and \figsref{fig:noRefinementInversions_p3_7}~to~\ref{fig:noRefinementInversions_p3_9} for $p = 3$. The quality of the reconstructions in this case depends mainly on the resolution of the material field. Increasing the number of voxels leads to a better reconstruction of the defect boundary. This, however, comes at the cost of a much higher computational effort, note \tabref{Tab:tab_p2} for polynomial degree $p=2$ and \tabref{Tab:tab_p3} for $p=3$, since the sensitivity kernel~\eqref{eq:sensitivity} has to be evaluated at every voxel of the material grid. For example, considering column $1$ (voxel size $1.25 \si{\milli \meter}$) of \tabref{Tab:tab_p2}, the computational time is dominated by solving the wave fields. In contrast, column $3$ (voxel size $0.3125\si{\milli \meter}$) uses the same approximation of the wave fields, but now spends most of the optimization effort evaluating the gradient. The time measurements clearly confirm the complexity estimation of equation~\eqref{eq:O_time}. Moreover, it can be seen that the reconstruction at undamaged regions is resolved quite well by a coarse voxel grid, since the reconstructed material does not vary largely. According to these observations, we suggest to use a locally refined material grid. The undamaged background can be resolved with a low number of voxels, while the areas of interest, i.e., where defects need to be detected, require a finer resolution. A corresponding refinement strategy and a suitable refinement indicator are presented in the following section.

\begin{table}[t]
	\centering
	\caption{Computation times for polynomial degree $p=2$}
	\renewcommand{\arraystretch}{1.5}
	\begin{tabular}{|l|c|c|c|}
		\hline
		& $v = 1.25 \si{\milli \meter}$ & $v = 0.625 \si{\milli \meter}$ & $v = 0.3125 \si{\milli \meter}$ \\
		\hline
		\hline $h = 5 \si{\milli \meter}$ & $\SI{158.4}{\second}$ & $\SI{460.7}{\second}$ & $\SI{1648.5}{\second}$ \\
		\hline
		$h = 2.5 \si{\milli \meter}$ & $\SI{283.4}{\second}$ & $\SI{590.7}{\second}$ & $\SI{2219.9}{\second}$ \\
		\hline
		$h = 1.25 \si{\milli \meter}$ & $\SI{620.2}{\second}$ & $\SI{1007.7}{\second}$ & $\SI{2556.7}{\second}$ \\
		\hline
	\end{tabular}
	\label{Tab:tab_p2}
\end{table}

\begin{table}[t]
	\centering
	\caption{Computation times for polynomial degree $p=3$}
	\renewcommand{\arraystretch}{1.5}
	\begin{tabular}{|l|c|c|c|}
		\hline
		& $v = 1.25 \si{\milli \meter}$ & $v = 0.625 \si{\milli \meter}$ & $v = 0.3125 \si{\milli \meter}$ \\
		\hline
		\hline $h = 5 \si{\milli \meter}$ & $\SI{268.4}{\second}$ & $\SI{790.5}{\second}$ & $\SI{2600.6}{\second}$ \\
		\hline
		$h = 2.5 \si{\milli \meter}$ & $\SI{425.1}{\second}$ & $\SI{965.4}{\second}$ & $\SI{3016.8}{\second}$ \\
		\hline
		$h = 1.25 \si{\milli \meter}$ & $\SI{990.8}{\second}$ & $\SI{1474.8}{\second}$ & $\SI{4004.5}{\second}$ \\
		\hline
	\end{tabular}
	\label{Tab:tab_p3}
\end{table}

\FloatBarrier
\subsection{Adaptive refinement of the material grid}
\FloatBarrier

For the refinement of the material grid, we introduce an indicator $\eta$ corresponding to each voxel with value $\hat{\gamma}_i$. Motivated by Sobel filters, which are used in image processing \cite{Gonzalez2017}, the $L2$-norm of the spatial gradient is used in \cite{Herrmann2023} for the sharpness quantification of a reconstructed material parameter, i.e.,
\begin{equation}
    \Vert \gamma(\tensor{x}) \Vert_2 = \sqrt{ \left( \frac{\partial \gamma(\tensor{x})}{\partial x} \right)^2 + \left( \frac{\partial  \gamma(\tensor{x})}{\partial y} \right)^2 }
    \label{sharpness}
\end{equation}
in two spatial dimensions.

High values indicate areas of rapidly changing material and, thus, boundaries of our regions of interest. Since the material parameter is discretized by constant shape functions defined on the voxel grid, we adapt the definition of the sharpness \eqref{sharpness} to introduce a suitable indicator, replacing the derivatives in the spatial directions by $G_x(\hat{\gamma}_i)$ and $G_y(\hat{\gamma}_i)$. This voxelized gradient is computed as the mean of the absolute jump values of the material parameter $\gamma$ between neighboring voxels, i.e.,
\begin{align}
    G_x(\hat{\gamma}_i) = \frac{1}{2 h^\text{v}}\left( \lvert \llbracket \hat{\gamma}_i \rrbracket^{(x,+)} \rvert + \lvert \llbracket \hat{\gamma}_i \rrbracket^{(x,-)} \rvert \right) = \frac{1}{2 h^\text{v}} \left( \lvert\hat{\gamma}_r - \hat{\gamma}_i \rvert + \lvert \hat{\gamma}_i - \hat{\gamma}_l \rvert \right) \\
    G_y(\hat{\gamma}_i) = \frac{1}{2 h^\text{v}}\left( \lvert \llbracket \hat{\gamma}_i \rrbracket^{(y,+)} \rvert + \lvert \llbracket \hat{\gamma}_i \rrbracket^{(y,-)} \rvert \right) = \frac{1}{2 h^\text{v}} \left( \lvert \hat{\gamma}_o - \hat{\gamma}_i \rvert + \lvert \hat{\gamma}_i - \hat{\gamma}_u \rvert \right)
\end{align}
where $h^\text{v}$ is the size of the voxel, $\hat{\gamma}_r$ is the voxel value of the voxel adjacent in positive $x$-direction and $\llbracket \hat{\gamma}_i \rrbracket^{(x,+)}$ is the jump of the material in that direction, $\hat{\gamma}_l$ and $\llbracket \hat{\gamma}_i \rrbracket^{(x,-)}$ are the voxel value and jump in negative $x$-direction. Equivalently, $\hat{\gamma}_o$, $\hat{\gamma}_u$, $\llbracket \hat{\gamma}_i \rrbracket^{(y,+)}$ and $\llbracket \hat{\gamma}_i \rrbracket^{(y,-)}$ are used in $y$-direction. Finally, the indicator $\eta$ of the voxel $i$ with value $\hat{\gamma}_i$ is evaluated as
\begin{equation}
    \eta(\hat{\gamma}_i) = \sqrt{ \left( G_x (\hat{\gamma}_i) \right)^2 + \left( G_y (\hat{\gamma}_i) \right)^2} \text{.}
\end{equation}

Using this indicator, we introduce an inversion framework with refinement. The basic ideas are shown in \figref{fig:conceptsRefinment}. It can readily be extended to 3D-problems.
\begin{figure}[H]
	\centering
	\begin{subfigure}[t]{0.3\textwidth}
		\centering
\begin{tikzpicture}

\begin{axis}[
axis equal,
hide axis,
xtick=\empty, 
ytick=\empty
]
\node at (0.3,1.5) {\small$\hat{\gamma}_l$};
\node at (1.5,1.5) {\small$\hat{\gamma}_i$};
\node at (2.7,1.5) {\small$\hat{\gamma}_r$};
\node at (1.5,0.3) {\small$\hat{\gamma}_u$};
\node at (1.5,2.7) {\small$\hat{\gamma}_o$};
\draw[->] (0.5,1.5)to[out=45,in=135](1.3,1.5);
\draw[->] (1.7,1.5)to[out=45,in=135](2.5,1.5);
\draw[->] (1.5,0.5)to[out=135,in=225](1.5,1.3);
\draw[->] (1.5,1.7)to[out=135,in=225](1.5,2.5);
\node at (0.8,1.85) {\small$\llbracket \hat{\gamma}_i \rrbracket^{(x,-)}$};
\node at (2.4,1.85) {\small$\llbracket \hat{\gamma}_i \rrbracket^{(x,+)}$};
\node at (2.05,0.8) {\small$\llbracket \hat{\gamma}_i \rrbracket^{(x,-)}$};
\node at (2.05,2.2) {\small$\llbracket \hat{\gamma}_i \rrbracket^{(x,+)}$};
\addplot [semithick, black]
table {%
0 0
1 0
1 1
0 1
0 0
};
\addplot [semithick, black]
table {%
1 0
2 0
2 1
1 1
1 0
};
\addplot [semithick, black]
table {%
2 0
3 0
3 1
2 1
2 0
};
\addplot [semithick, black]
table {%
0 1
1 1
1 2
0 2
0 1
};
\addplot [semithick, black]
table {%
1 1
2 1
2 2
1 2
1 1
};
\addplot [semithick, black]
table {%
2 1
3 1
3 2
2 2
2 1
};
\addplot [semithick, black]
table {%
0 2
1 2
1 3
0 3
0 2
};
\addplot [semithick, black]
table {%
1 2
2 2
2 3
1 3
1 2
};
\addplot [semithick, black]
table {%
2 2
3 2
3 3
2 3
2 2
};
\end{axis}

\end{tikzpicture}
		\caption{Jumps in the voxelized material parameter}
		\label{fig:computationJumps}
	\end{subfigure}
	\hspace{0.03\textwidth}
	\begin{subfigure}[t]{0.3\textwidth}
		\centering
\begin{tikzpicture}

\begin{axis}[
axis equal,
hide axis,
xtick=\empty, 
ytick=\empty
]
\addplot [semithick, black, fill=blue, fill opacity=0.5]
table {%
0 0
1 0
1 1
0 1
0 0
};
\addplot [semithick, black, fill=red, fill opacity=0.2]
table {%
1 0
2 0
2 1
1 1
1 0
};
\addplot [semithick, black]
table {%
2 0
3 0
3 1
2 1
2 0
};
\addplot [semithick, black]
table {%
3 0
4 0
4 1
3 1
3 0
};
\addplot [semithick, black]
table {%
4 0
5 0
5 1
4 1
4 0
};
\addplot [semithick, black]
table {%
5 0
6 0
6 1
5 1
5 0
};
\addplot [semithick, black]
table {%
6 0
7 0
7 1
6 1
6 0
};
\addplot [semithick, black]
table {%
7 0
8 0
8 1
7 1
7 0
};
\addplot [semithick, black, fill=red, fill opacity=0.2]
table {%
0 1
1 1
1 2
0 2
0 1
};
\addplot [semithick, black, fill=red, fill opacity=0.2]
table {%
1 1
2 1
2 2
1 2
1 1
};
\addplot [semithick, black]
table {%
2 1
3 1
3 2
2 2
2 1
};
\addplot [semithick, black]
table {%
3 1
4 1
4 2
3 2
3 1
};
\addplot [semithick, black]
table {%
4 1
5 1
5 2
4 2
4 1
};
\addplot [semithick, black]
table {%
5 1
6 1
6 2
5 2
5 1
};
\addplot [semithick, black]
table {%
6 1
7 1
7 2
6 2
6 1
};
\addplot [semithick, black]
table {%
7 1
8 1
8 2
7 2
7 1
};
\addplot [semithick, black]
table {%
0 2
1 2
1 3
0 3
0 2
};
\addplot [semithick, black]
table {%
1 2
2 2
2 3
1 3
1 2
};
\addplot [semithick, black]
table {%
2 2
3 2
3 3
2 3
2 2
};
\addplot [semithick, black]
table {%
3 2
4 2
4 3
3 3
3 2
};
\addplot [semithick, black]
table {%
4 2
5 2
5 3
4 3
4 2
};
\addplot [semithick, black]
table {%
5 2
6 2
6 3
5 3
5 2
};
\addplot [semithick, black]
table {%
6 2
7 2
7 3
6 3
6 2
};
\addplot [semithick, black]
table {%
7 2
8 2
8 3
7 3
7 2
};
\addplot [semithick, black]
table {%
0 3
1 3
1 4
0 4
0 3
};
\addplot [semithick, black]
table {%
1 3
2 3
2 4
1 4
1 3
};
\addplot [semithick, black]
table {%
2 3
3 3
3 4
2 4
2 3
};
\addplot [semithick, black]
table {%
3 3
4 3
4 4
3 4
3 3
};
\addplot [semithick, black, fill=red, fill opacity=0.2]
table {%
4 3
5 3
5 4
4 4
4 3
};
\addplot [semithick, black, fill=red, fill opacity=0.2]
table {%
5 3
6 3
6 4
5 4
5 3
};
\addplot [semithick, black, fill=red, fill opacity=0.2]
table {%
6 3
7 3
7 4
6 4
6 3
};
\addplot [semithick, black]
table {%
7 3
8 3
8 4
7 4
7 3
};
\addplot [semithick, black]
table {%
0 4
1 4
1 5
0 5
0 4
};
\addplot [semithick, black]
table {%
1 4
2 4
2 5
1 5
1 4
};
\addplot [semithick, black]
table {%
2 4
3 4
3 5
2 5
2 4
};
\addplot [semithick, black]
table {%
3 4
4 4
4 5
3 5
3 4
};
\addplot [semithick, black, fill=red, fill opacity=0.2]
table {%
4 4
5 4
5 5
4 5
4 4
};
\addplot [semithick, black, fill=blue, fill opacity=0.5]
table {%
5 4
6 4
6 5
5 5
5 4
};
\addplot [semithick, black, fill=red, fill opacity=0.2]
table {%
6 4
7 4
7 5
6 5
6 4
};
\addplot [semithick, black]
table {%
7 4
8 4
8 5
7 5
7 4
};
\addplot [semithick, black]
table {%
0 5
1 5
1 6
0 6
0 5
};
\addplot [semithick, black]
table {%
1 5
2 5
2 6
1 6
1 5
};
\addplot [semithick, black]
table {%
2 5
3 5
3 6
2 6
2 5
};
\addplot [semithick, black]
table {%
3 5
4 5
4 6
3 6
3 5
};
\addplot [semithick, black, fill=red, fill opacity=0.2]
table {%
4 5
5 5
5 6
4 6
4 5
};
\addplot [semithick, black, fill=blue, fill opacity=0.5]
table {%
5 5
6 5
6 6
5 6
5 5
};
\addplot [semithick, black, fill=red, fill opacity=0.2]
table {%
6 5
7 5
7 6
6 6
6 5
};
\addplot [semithick, black]
table {%
7 5
8 5
8 6
7 6
7 5
};
\addplot [semithick, black]
table {%
0 6
1 6
1 7
0 7
0 6
};
\addplot [semithick, black]
table {%
1 6
2 6
2 7
1 7
1 6
};
\addplot [semithick, black]
table {%
2 6
3 6
3 7
2 7
2 6
};
\addplot [semithick, black]
table {%
3 6
4 6
4 7
3 7
3 6
};
\addplot [semithick, black, fill=red, fill opacity=0.2]
table {%
4 6
5 6
5 7
4 7
4 6
};
\addplot [semithick, black, fill=red, fill opacity=0.2]
table {%
5 6
6 6
6 7
5 7
5 6
};
\addplot [semithick, black, fill=red, fill opacity=0.2]
table {%
6 6
7 6
7 7
6 7
6 6
};
\addplot [semithick, black]
table {%
7 6
8 6
8 7
7 7
7 6
};
\addplot [semithick, black]
table {%
0 7
1 7
1 8
0 8
0 7
};
\addplot [semithick, black]
table {%
1 7
2 7
2 8
1 8
1 7
};
\addplot [semithick, black]
table {%
2 7
3 7
3 8
2 8
2 7
};
\addplot [semithick, black]
table {%
3 7
4 7
4 8
3 8
3 7
};
\addplot [semithick, black]
table {%
4 7
5 7
5 8
4 8
4 7
};
\addplot [semithick, black]
table {%
5 7
6 7
6 8
5 8
5 7
};
\addplot [semithick, black]
table {%
6 7
7 7
7 8
6 8
6 7
};
\addplot [semithick, black]
table {%
7 7
8 7
8 8
7 8
7 7
};
\end{axis}

\end{tikzpicture}
		\caption{Indicated voxels (blue) and one surrounding layer (red)}
		\label{fig:indicatedAreas}
	\end{subfigure}
	\hspace{0.03\textwidth}
	\begin{subfigure}[t]{0.3\textwidth}
		\centering
		\input{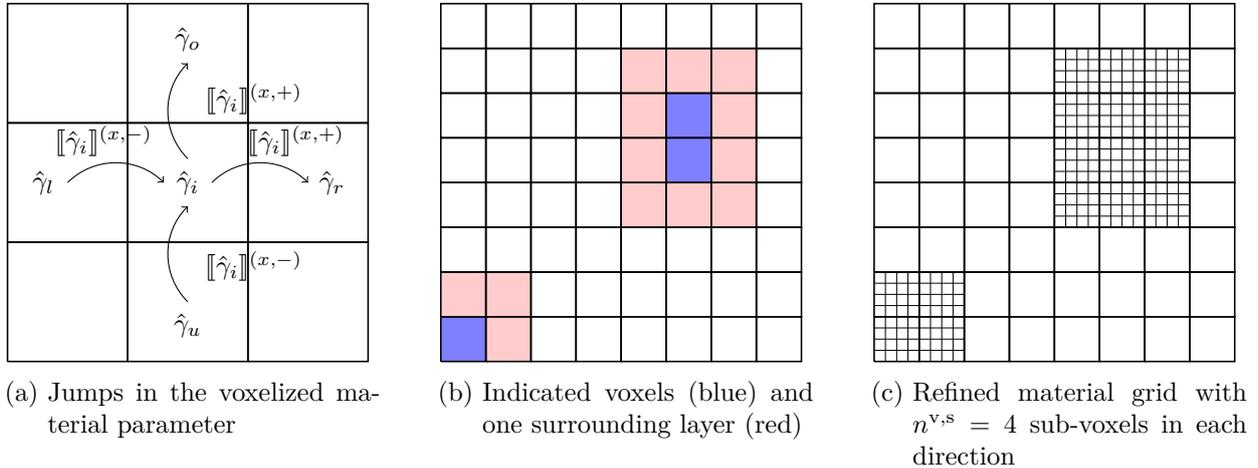}
		\caption{Refined material grid with $n^\text{v,s}=4$ sub-voxels in each direction}
		\label{fig:refinedAreas}
	\end{subfigure}
	\caption{Refinement of the material grid}
	\label{fig:conceptsRefinment}
\end{figure}
The reconstruction of the material parameter starts with an inversion on a coarse material grid with $n^\text{v}$ voxels per knot span in each spatial direction. After $N^{i,1}$ iterations, the indicator is evaluated for each voxel of this intermediate solution and then compared to a threshold value $\tau$. In the following examples, this threshold is set to half of the maximum occurring indicator value
\begin{equation}
    \tau = \frac{1}{2} \arg \max_i \eta(\hat{\gamma}_i) \text{.}
\end{equation}
Since the material interfaces may not be perfectly identified yet, the indicated voxels and $n^\text{l}$ surrounding layers are refined into $n^\text{v,s}$ sub-voxels per spatial direction. The parameter coefficients of the constant shape functions corresponding to these sub-voxels are included in the set of optimization variables. Finally, $N^{i,2}$ iterations are performed in a second inversion. 

As in the previous section the example of \figref{embeddedDomain_withFlaw} is considered. Locally refined inversions are performed for a wave field discretization of $h = 1.25 \si{\milli \meter}$ and polynomial degree $p = 2$ or $p = 3$. For the first $N^{i,1} = 3$ iterations, only one voxel per knot span is used to model the material, i.e., $n^\text{v} = 1$. This intermediate solution identifies the area of interest to be locally refined. The indicated voxels and one additional surrounding layer are subdivided into $n^\text{v,s} = 4$ sub-voxels in each spatial direction. From here, two different variants are investigated. In the first one, the intermediate solution of the first three iterations is chosen as the initial guess for the following $N^{i,2} = 7$ iterations. In the second variant, a full restart, the inversion is again performed for $N^{i,2} = 10$ iterations starting from homogeneous material. \figref{RefinementInversions_p2} shows the results of the inversion with the introduced local refinement strategies for $p = 2$ and \figref{RefinementInversions_p3} for $p = 3$. The intermediate results from the first three iterations, the corresponding sharpness and refined areas, and the final inversion results for both strategies are depicted. It is obvious that the reconstruction quality of the restart variant is superior to that of the start with an initial guess. The reason for this is that the intermediate reconstruction can be already trapped in a local minimum of the optimization process, which can yet not be refined to a minimum on the finer grid.

\begin{figure}[H]
	\centering
	\begin{subfigure}[t]{0.3\textwidth}
		\centering
		\includegraphics[width=\textwidth]{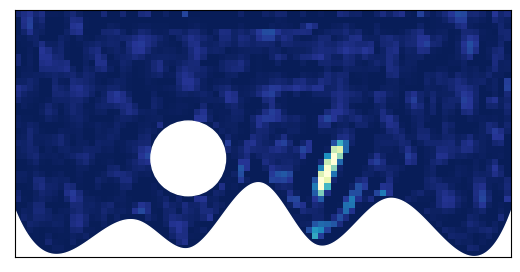}
		\caption{Iteration $3$}
	\end{subfigure}
	\begin{subfigure}[t]{0.3\textwidth}
		\centering
		\includegraphics[width=\textwidth]{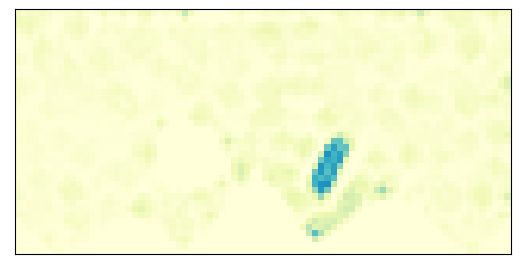}
		\caption{Sharpness}
	\end{subfigure}
	\begin{subfigure}[t]{0.3\textwidth}
		\centering
		\includegraphics[width=\textwidth]{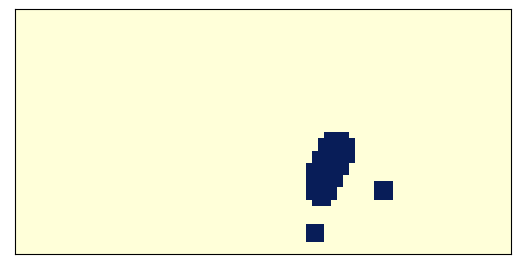}
		\caption{Refinement}
	\end{subfigure}
	\begin{subfigure}[t]{0.45\textwidth}
		\centering
		\input{tikzpictures/final_p2_v1_withZoom.tex}
		\caption{Iteration $10$ -- start with initial guess}
	\end{subfigure}
	\begin{subfigure}[t]{0.45\textwidth}
		\centering
		\input{tikzpictures/final_p2_v2_withZoom.tex}
		\caption{Iteration $10$ -- restart version}
	\end{subfigure}
	\caption{Inversion results with refinement -- $h = 1.25 \si{\milli \meter}$ and $p=2$}
	\label{RefinementInversions_p2}
\end{figure}

\begin{figure}[H]
	\centering
	\begin{subfigure}[t]{0.3\textwidth}
		\centering
		\includegraphics[width=\textwidth]{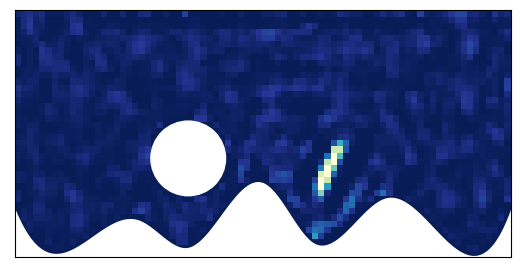}
		\caption{Iteration $3$}
	\end{subfigure}
	\begin{subfigure}[t]{0.3\textwidth}
		\centering
		\includegraphics[width=\textwidth]{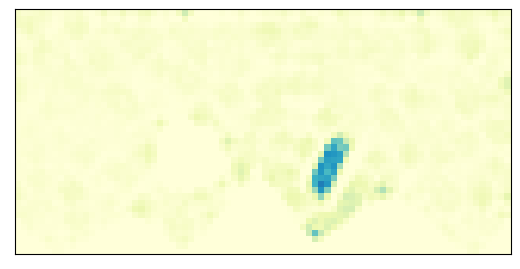}
		\caption{Sharpness}
	\end{subfigure}
	\begin{subfigure}[t]{0.3\textwidth}
		\centering
		\includegraphics[width=\textwidth]{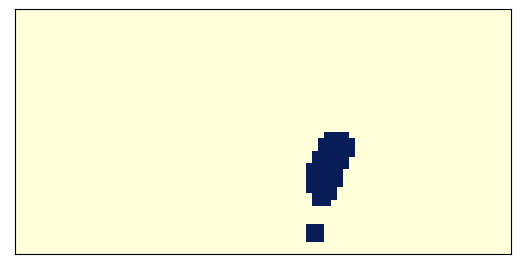}
		\caption{Refinement}
	\end{subfigure}
	\begin{subfigure}[t]{0.45\textwidth}
		\centering
		\input{tikzpictures/final_p3_v1_withZoom.tex}
		\caption{Iteration $10$ -- start with initial guess}
	\end{subfigure}
	\begin{subfigure}[t]{0.45\textwidth}
		\centering
		\input{tikzpictures/final_p3_v2_withZoom.tex}
		\caption{Iteration $10$ -- restart version}
	\end{subfigure}
	\caption{Inversion results with refinement -- $h = 1.25 \si{\milli \meter}$ and $p=3$}
	\label{RefinementInversions_p3}
\end{figure}

\tabref{Tab:tab_refined} shows the computational times of the investigated refinement strategies. While the effort for the restart version is moderately larger than that of the refinement using the initial guess of the coarse material grid, both locally refined variants are over two times faster than the unrefined inversion. This is due to the fact that the computational effort for the gradient computation is greatly reduced. It should be noted that the restart version, in particular, does not compromise the quality of the reconstruction. 

\begin{table}[H]
	\centering
	\caption{Computation times with refinement for $h = 1.25 \si{\milli \meter}$}
	\renewcommand{\arraystretch}{1.5}
	\begin{tabular}{|l|c|c|}
	    \hline
	    & $p=2$ & $p=3$ \\
	    \hline
	    \hline
	    no refinement & $\SI{2556.7}{\second}$ & $\SI{4004.5}{\second}$ \\
	    \hline
	    refinement -- start with initial guess & $\SI{945.8}{\second}$ & $\SI{1328.8}{\second}$ \\
	    \hline
	    refinement -- restart version & $\SI{1019.1}{\second}$ & $\SI{1813.5}{\second}$ \\
	    \hline
	\end{tabular}
	\label{Tab:tab_refined}
\end{table}

\FloatBarrier
\subsection{3D example} \label{sec:3Dexample}
\FloatBarrier

The 3D structure under consideration is shown in \figref{3Dstructure}. The left pillar with the corresponding part of the roof (marked in blue) of the structure is examined locally in two inversions. The surface is given in STL (`Standard Triangle Language') format. At first only the left pillar is embedded in an extended computational domain of size $\SI{2}{\meter} \times \SI{1.25}{\meter} \times \SI{1.25}{\meter}$. The density is set to $\SI{2400}{\kilogram \per \meter^3}$, the wave speed to $\SI{3000}{\meter \per \second}$. The setup of the inversion is shown in \figref{pillarReference}. 

\begin{figure}[H]
	\centering
	\includegraphics[width=0.7\textwidth]{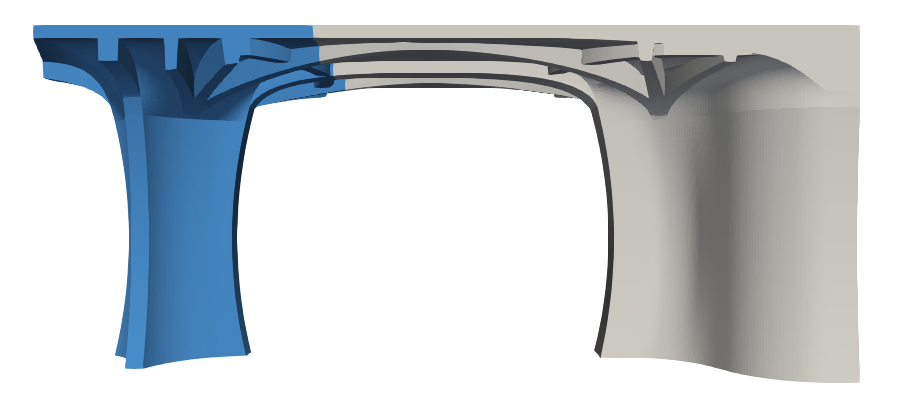}
	\caption{Investigated 3D structure~\cite{kloft2023}}
	\label{3Dstructure}
\end{figure}

\begin{figure}[H]
	\centering
	\begin{subfigure}[t]{0.3\textwidth}
		\centering
		\includegraphics[width=0.8\textwidth]{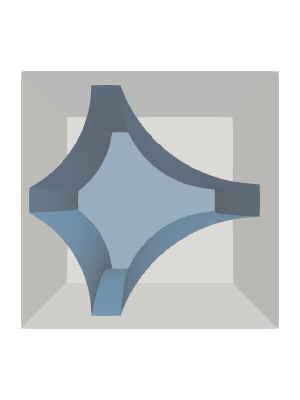}
		\includegraphics[width=0.8\textwidth]{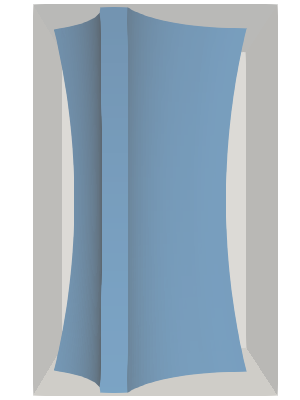}
		\caption{Computational domain}
	\end{subfigure}
	\begin{subfigure}[t]{0.3\textwidth}
		\centering
		\includegraphics[width=0.8\textwidth]{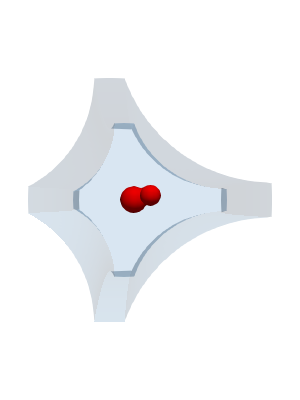}
		\includegraphics[width=0.8\textwidth]{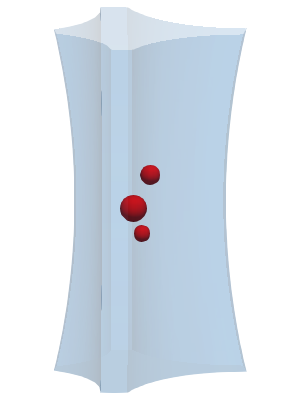}
		\caption{Cavities}
	\end{subfigure}
	\begin{subfigure}[t]{0.3\textwidth}
		\centering
		\includegraphics[width=0.8\textwidth]{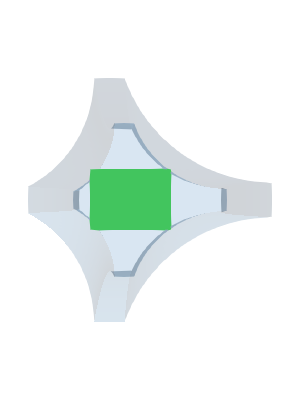}
		\includegraphics[width=0.8\textwidth]{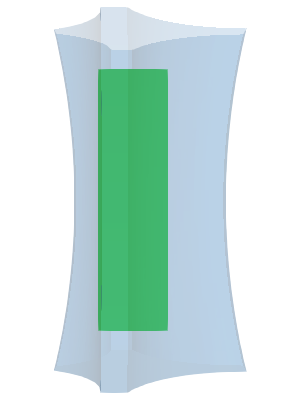}
		\caption{Search window}
	\end{subfigure}
	\caption{FWI setup of the left pillar: The structure is colored in blue, the computational domain in gray, the circular cavities in red, and the search window in green. Top views are given in the top pictures, while front views are displayed in the bottom pictures.}
	\label{pillarReference}
\end{figure}

Three cavities are introduced at heights $\SI{0.8}{\meter}$, $\SI{0.95}{\meter}$, and $\SI{1.15}{\meter}$ with radii $\SI{0.05}{\meter}$, $\SI{0.08}{\meter}$, and $\SI{0.06}{\meter}$. Reference data are generated for twelve sources placed in three different heights, i.e., $\SI{0.5}{\meter}$, $\SI{1.0}{\meter}$, and $\SI{1.5}{\meter}$, using a mesh of quadratic B-splines with knot span size $h = \SI{0.025}{\meter}$. Within the void regions, $\alpha$ is set to $10^{-6}$. Integration of the system matrices is performed with an octree of depth $3$. The source term is a 2-cycle sine burst with a central frequency $f = \SI{10}{\kilo \hertz}$, resulting in a dominant wave length $\lambda_\text{dom} = \SI{0.3}{\meter}$. In the inversion, the wave fields are discretized by cubic B-splines defined on a mesh with knot span size $h = \SI{0.05}{\meter}$. The simulation of a time span of $\SI{8.0 e-4}{\second}$ is carried out in $800$ time integration steps. The material field is first defined on a grid with $n^\text{v} = 2$ voxels per knot span in each direction, before it is locally refined after $N^{i,1} = 3$ iterations with $n^\text{v,s} = 4$ sub-voxels per voxel in each direction. Integration of the system matrices is carried out on the voxel-level, incorporating the a priori known geometry by an octree of depth $4$ for the cut knot spans with $\alpha = 10^{-5}$. Both local refinement strategies are executed -- with either $N^{i,2} = 7$ additional iterations, if the intermediate solution is chosen as initial model for the subsequent optimization, or $N^{i,2} = 10$, if a homogeneous material is chosen as the new initial model. FWI can be applied to a region of interest by introducing a search window, see e.g., \cite{Rabinovich2023}. Consequently, $\gamma$ is just optimized within the selected region. In this example, the search window has a size of $\SI{0.4}{\meter} \times \SI{0.3}{\meter} \times \SI{1.5}{\meter}$ and is centered at $\SI{1}{\meter}$ height, mimicking the inner of the pillar. The optimization is bounded between $\gamma_\text{min} = 10^{-5}$ and $\gamma_\text{max} = 1$. The reconstructed cavities are shown in \figref{pillarResults}. For a better visualization, the Iso Volume filter of ParaView was applied to the voxelized representation of the defects. The threshold in $\gamma$ is set to $0.5$ to classify void regions. Both strategies are suitable to precisely identify the positions and sizes of the defects. Also, the spherical shape of the voids is accurately approximated. In the restart version, the surface is reconstructed smoother due to the fact that the inversion is not trapped in a local minimum caused by the coarse material grid.

\begin{figure}[H]
	\centering
	\begin{subfigure}[t]{0.3\textwidth}
		\centering
		\includegraphics[width=\textwidth]{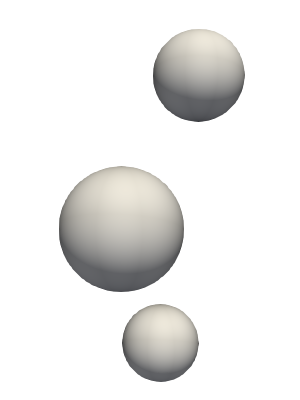}
		\caption{Reference solution}
	\end{subfigure}
	\begin{subfigure}[t]{0.3\textwidth}
		\centering
		\includegraphics[width=\textwidth]{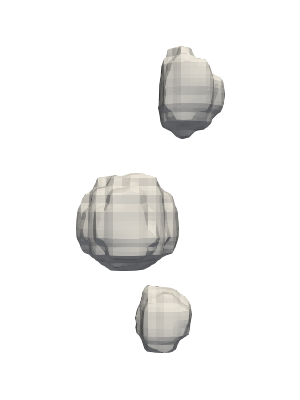}
		\caption{Refined inversion \\-- start with initial guess}
	\end{subfigure}
	\begin{subfigure}[t]{0.3\textwidth}
		\centering
		\includegraphics[width=\textwidth]{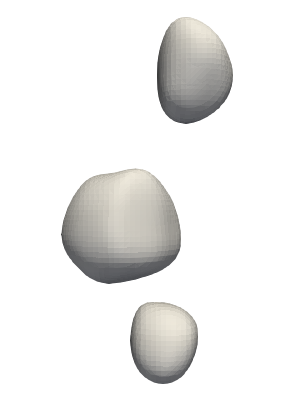}
		\caption{Refined inversion \\-- restart version}
	\end{subfigure}
	\caption{Reconstructed cavities in the left pillar}
	\label{pillarResults}
\end{figure}

In a second inversion, the geometrically more complex left roof part is investigated. The FWI setup is shown in~\figref{roofReference}. The physical domain is embedded in a computational domain of size $\SI{2.5}{\meter} \times \SI{2.5}{\meter} \times \SI{0.85}{\meter}$. Nine sources are positioned on the top surface, while six sources are located on the bottom surface. To improve the inversion process, a search window is defined, excluding the areas where the sources are attached. In the reference model, an ellipsoidal cavity is placed at position $\left( \SI{1.35}{\meter}, \SI{1.2}{\meter}, \SI{0.6}{\meter}\right)$ with semi-axes $\SI{0.15}{\meter}$, $\SI{0.075}{\meter}$, and $\SI{0.075}{\meter}$. The material parameters, spatial and time discretizations, and the source term remain the same as for the previous pillar example. Both refinement strategies are performed in the same manner as in the previous example. \figref{roofResults} shows the identified cavities. In both strategies, the ellipsoid is detected at the right position with a proper shape. It has to be noted that the strategy that continues with the intermediate solution terminates after three refined iterations. The optimization is trapped in a local minimum and is not able to find a suitable update. Consequently, the size of the defect is slightly underestimated. On the other hand, the restart version successfully reproduces position, shape, and size of the cavity.

\FloatBarrier
\begin{figure}[H]
	\centering
	\begin{subfigure}[t]{0.45\textwidth}
		\centering
		\includegraphics[width=1.0\textwidth]{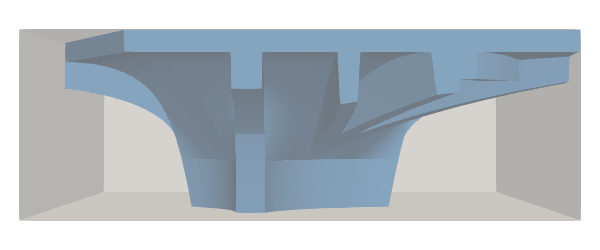}
		\caption{Computational domain}
	\end{subfigure}
	\begin{subfigure}[t]{0.45\textwidth}
		\centering
		\includegraphics[width=1.0\textwidth]{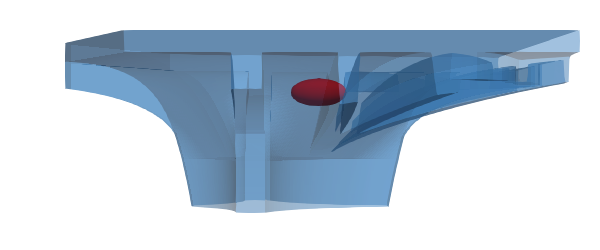}
		\caption{Cavity}
	\end{subfigure}
	\begin{subfigure}[t]{0.45\textwidth}
		\centering
		\includegraphics[width=1.0\textwidth]{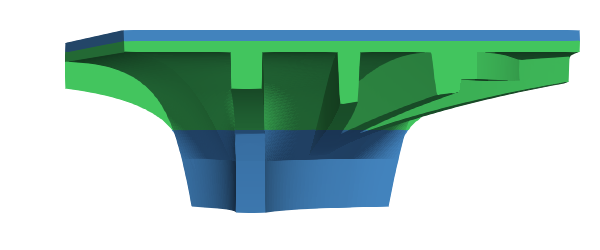}
		\caption{Search window}
	\end{subfigure}
	\caption{FWI setup of the left roof part: The structure is colored in blue, the computational domain in gray, the circular cavities in red, and the search window in green. Top views are given in the top pictures, while front views are displayed in the bottom pictures.}
	\label{roofReference}
\end{figure}

\begin{figure}[H]
	\centering
	\begin{subfigure}[t]{0.3\textwidth}
		\centering
		\includegraphics[width=0.8\textwidth]{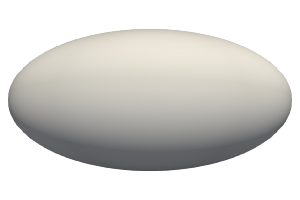}
		\caption{Reference solution}
	\end{subfigure}
	\begin{subfigure}[t]{0.3\textwidth}
		\centering
		\includegraphics[width=0.8\textwidth]{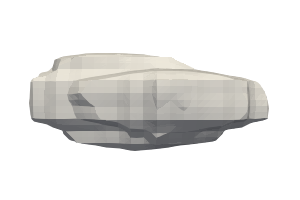}
		\caption{Refined inversion \\-- start with initial guess}
	\end{subfigure}
	\begin{subfigure}[t]{0.3\textwidth}
		\centering
		\includegraphics[width=0.8\textwidth]{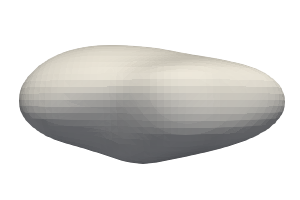}
		\caption{Refined inversion \\-- restart version}
	\end{subfigure}
	\caption{Reconstructed cavities in the roof}
	\label{roofResults}
\end{figure}

\FloatBarrier
}
\section{Conclusion}
{ 
\label{sec5}
\FloatBarrier

In the paper at hand, we propose a multi-resolution FWI approach based on a IGA-FCM discretization of the wave field and a voxelized representation of the material. In \secref{sec3}, an introductory investigation of the forward problem shows that IGA-FCM, used with a consistent mass matrix, is well-suited to solve the scalar wave equation precisely with a high accuracy per degree of freedom. Considering the inverse problem (\secref{sec4}), if the wave field is adequately resolved, the quality of the reconstruction mainly depends on the representation of the material field. By increasing the resolution of the independent material representation, a more precise identification of the defect boundaries is possible. However, this comes at the cost of rapidly increasing computation times, since the gradient has to be evaluated at each voxel. To mitigate this computational burden, we suggest a method to locally refine the material based on an indicator that accounts for the local changes of the material. The inversion is decomposed into a two-step optimization. At first, one optimization is performed on a coarse material grid. The resulting intermediate solution is then used to indicate regions of interest where the material field is refined. Finally, a second optimization is carried out on the locally refined material grid. The intermediate solution can serve as an initial model -- or, alternatively, a restart is carried out starting with a homogeneous initial material. Particularly, the restart strategy leads to an accurate reconstruction of the defect's location, size, and shape, despite coming at a slightly higher but still reasonable computational cost. Finally, the multi-resolution approach using local refinement is applied to a 3D specimen. Spherical and ellipsoidal cavities are identified and quantified accurately within a few iterations. 

In summary, the paper at hand provides a framework that enables accurate and efficient FWI for the detection of void regions validated for synthetic reference data. IGA-FCM is used to discretize the wave field with a low number of degrees of freedom. The proposed indicator and refinement strategy of the material enables a precise and effective reconstruction of the defects. Our future research will further explore the potential and limits of this approach, in particular when it is applied to experimental data sets. Therefore, we plan to extend the framework to the elastic wave equation.

\FloatBarrier
}

\appendix


\section*{Acknowledgements} \label{sec:Acknowledgement}
{
	We gratefully thank the Deutsche Forschungsgemeinschaft (DFG, German Research Foundation) for their support through the grants KO 4570/1-1 and RA 624/29-1. We also thank Harald Kloft and Robin D\"{o}rrie from the Institut f\"{u}r Tragwerksentwurf of the Technische Universit\"{a}t Braunschweig for providing the geometric model investigated in~\secref{sec:3Dexample} which serves as a benchmark in the DFG project 414265976 TRR 277.
}


\bibliographystyle{ieeetr}

\setlength{\bibsep}{3pt}
\setlength{\bibhang}{0.75cm}{\fontsize{9}{9}\selectfont\bibliography{library}}

\end{document}